\documentclass{ptephy_v2}%%%%%% to generate preprint number with ptep logo

\usepackage{amsmath}
\usepackage{graphics}
\usepackage{multirow}
\usepackage{url}
\usepackage{float}
\usepackage{color}
\begin{document}
 
 % \title{Gamma Production from Thermal Neutron Capture on Gadolinium-157}
\title{Gamma Ray Spectra from Thermal Neutron Capture on Gadolinium-155 and Natural Gadolinium}

%%%% To generate auto affiliation numbers please use \author{}\affil{} command

\author[1]{Tomoyuki Tanaka}
\author[1]{Kaito Hagiwara}
\author[3]{Enrico Gazzola}
\author[1,6,*]{Ajmi~Ali}
\author[1]{Iwa Ou}
\author[1,8]{Takashi Sudo}
\author[1,9]{Pretam Kumar Das}
\author[1,10]{Mandeep Singh Reen}
\author[1,11]{Rohit Dhir}
\author[1]{Yusuke Koshio}
\author[1,*]{Makoto Sakuda}
\author[2]{Atsushi Kimura}
\author[2]{Shoji Nakamura}
\author[2]{Nobuyuki Iwamoto}
\author[2]{Hideo Harada}
\author[3]{Gianmaria Collazuol}
\author[4]{Sebastian Lorenz}
\author[4]{Michael Wurm}
\author[5]{William Focillon}
\author[5]{Michel Gonin}
\author[7]{Takatomi Yano}

\affil[1]{Department of Physics, Okayama University, Okayama 700-8530, Japan }
\affil[2]{Japan Atomic Energy Agency, 2-4 Shirakata Shirane, Tokai, Naka, Ibaraki 319-1195, Japan}
\affil[3]{Universit\'a di Padova and INFN, Dipartimento di Fisica, Padova 35131, Italy}
\affil[4]{Institut f\"ur Physik, Johannes Gutenberg-Universit\"at Mainz, 55128 Mainz, Germany}
\affil[5]{D\'epartement de Physique, \'Ecole Polytechnique, 91128 Palaiseau Cedex, France}
\affil[6]{Present address: Department of Physics, Kyoto University, Kyoto 606-8502, Japan}
\affil[7]{Present address: Kamioka Observatory, ICRR, University of Tokyo, Gifu 506-1205, JAPAN}
\affil[8]{Present address: Research Center for Nuclear Physics (RCNP),
Osaka University,  Osaka 567-0047, Japan}
\affil[9]{Present address: Department of Physics, Pabna University of Science $\&$ Technology, 
Pabna-6600, Bangladesh}
\affil[10]{Present Address: Department of Physics, Akal University, Punjab 151302, India}
\affil[11]{Present address: Research Institute $\&$ Department of Physics and Nano Technology, SRM 
University, Kattankulathur-603203,  India}
\affil[ ]{\email{ali.ajmi.3c@kyoto-u.ac.jp, sakuda-m@okayama-u.ac.jp}}

\begin{abstract}%
% We measured the $\gamma$-ray energy spectrum from the radiative thermal neutron capture, 
% ${}^{155}$Gd$(n,\gamma){}^{156}$Gd, on an enriched $^{155}$Gd target (Gd$_{2}$O$_{3}$) in the 
% energy range from 0.11 MeV up to about 8 MeV. The target was placed inside the germanium 
% spectrometer of the ANNRI detector at MLF, J-PARC and exposed to a neutron beam from the Japan 
% Spallation Neutron Source. 
% Radioactive sources ($^{60}$Co, $^{137}$Cs, and $^{152}$Eu) and the 
% reaction $^{35}$Cl($n$,$\gamma$) were used to determine the spectrometer's detection efficiency for 
% $\gamma$-rays at energies from 0.3 to 8.5 MeV. Using 

Natural gadolinium is widely used for its excellent thermal neutron capture cross section, because of its two major isotopes: $^{\rm 155}$Gd and $^{\rm 157}$Gd.
  We measured the $\gamma$-ray spectra produced from the thermal neutron capture on targets comprising a natural gadolinium film and enriched $^{\rm 155}$Gd  (in Gd$_{2}$O$_{3}$ powder) in the energy range from 0.11 MeV to 8.0 MeV, using the ANNRI germanium spectrometer at MLF, J-PARC. 
  The freshly analysed data of the $^{\rm 155}$Gd(n, $\gamma$) reaction are used to improve our previously developed model (ANNRI-Gd model)
for the   $^{\rm 157}$Gd(n, $\gamma$)  reaction~\cite{Hagiwara:2018kmr}, and its performance confirmed with the independent data from the $^{\rm nat}$Gd(n, $\gamma$) reaction. This article completes the development of an efficient Monte Carlo model required to  simulate and analyse particle interactions involving the thermal neutron captures on gadolinium in any relevant future experiments. 
  
\end{abstract}

\subjectindex{D21 Models of nuclear reactions, F22 Neutrinos from supernova and other astronomical objects, C43 Underground experiments, F20 Instrumentation and technique, H20 Instrumentation for underground experiments, H43 Software architectures
}

\maketitle

\newpage
%==================================================================================================%
\section{Introduction}
%==================================================================================================%

Gadolinium (Gd) 
has become an important element of consideration to a number of neutrino experiments 
 for enhanced detection of electron anti-neutrinos ($\bar{\nu}_e$). The presence of Gd boosts the tagging of neutrons in the inverse beta decay reaction (IBD), $\bar\nu_e + p\ \rightarrow \ e^++ n $, in organic liquid scintillator and water-Cherenkov detectors.  This is primarily due to its large capture cross-section for thermal neutrons and the large energy released by  $\gamma$ rays of $\sim8 {\rm MeV}$ for the Gd$(n,\gamma)$) reactions~\cite{Mughabghab2006:NeutRes, Leinweber, Choi2007:EGAF, nTOF},
%  \footnote{The coincidence $\gamma$ rays are emitted on neutron capture  shortly after ($\sim$20$\mu$s in water) the prompt  signal from the annihilation of the $e^+$ in the IBD. We wrote in PTEP:The mean timescale 
% $\tau_{\mathrm{cap}}$ for the neutron capture depends on the concentrations $n_i$ and the thermal 
% neutron capture cross-sections $\sigma_{\mathrm{cap},i}$ of the nuclei $i$  in the detector 
% material as well as on the mean velocity $v_n$ of the produced neutrons: 
% $\tau_{\mathrm{cap}} \propto 1/(n_i \, \sigma_{\mathrm{cap},i} \, v_n)$. The time scale depends on the concentration, density, cross section. It is not 20 microsec when the concentration is not 0.1\%.  Either you give a typical time sce for liq.scintillator and water. } 
%I do not think that \footnote{See \cite{Hagiwara:2018kmr} for details.} is necessary any more. 
% listed in Table~\ref{tbl:CommonCatcherIsotopes}. 

{\centering
$
 {\rm n} + ^{155}{\rm Gd} \rightarrow ^{156}{\rm Gd}^\ast \rightarrow ^{156}{\rm Gd} + {\bf \gamma}\ {\rm {rays}\ (8.536\ MeV\ total)}
$, and

$
 {\rm n} + ^{157}{\rm Gd} \rightarrow ^{158}{\rm Gd}^\ast \rightarrow ^{158}{\rm Gd} + {\bf \gamma}\ {\rm {rays}\ (7.937\ MeV\ total)}.
$

}

 The element has already been used as a neutron absorber in
scintillator-based detectors for the neutrino oscillation experiments~\cite{Abe2012:DC_NuEbarDisapp, Ahn2012:RENO_NuEbarDisapp, An2012:DB_NuEbarDisapp, Ko:2016owz, STEREO, DANSS, Neutrino-4, JSNS2} 
and a neutrino-flux monitor experiment ~\cite{Oguri2014:PANDA}. 
 For the upcoming SuperKamiokande-Gd (SK-Gd) phase~\cite{Vagins, Sekiya2016:SkGd, Watanabe2009:NeutTagWCGd}, Gd will be dissolved in a multi-kiloton water-Cherenkov detector.  
 The application of  Gd-loaded detector materials for neutron tagging is foreseen for Direct Dark Matter Search experiments like LZ~\cite{Akerib:2019sek}   and XENONnT~\cite{Moriyama:2019}.  
%  Y.J.Ko et al.(NEOS), Phys.Rev.Lett.118, 121802(2017). 
%  D.S.Akerib et al.(LZ collab), arXiv:1904.02112, Fermilab-pub-19-139-AE-PPPD. 
%  S.Moriyama (for XENONnT collaboration), Direct Dark Matter Serach with XENONnT,
% presented in The International Symposium on Revealing the history of the Universe with Underground Particle and Nuclear Research,
% March 8, 2019, Tohoku University.
 
 Therefore,  it is of paramount importance to establish a precise Monte Carlo (MC) model for the $\gamma$-ray energy spectrum from the radiative thermal neutron capture on Gd.
%  is specially increased in the present era of Gd-enhanced $\bar{\nu}_e$-search detectors.  
 It is an essential prerequisite for MC studies aiming to evaluate the neutron tagging efficiency in a Gd-loaded detector. Precise modeling  is especially important for those detectors which lack hermetic acceptance or/and have a high energy threshold for $\gamma$-rays, since some of $\gamma$ rays emitted in the capture reaction may not be detected. 
 
    In most cases, detector materials are doped with the natural Gd ($^{\rm nat}$Gd).  Isotopic adundances are listed in  Table~\ref{tab:gdisotopes}. 
\begin{table}[H]
\caption{Relative abundances of gadolinium isotopes in natural 
             gadolinium~\cite{Rosman1998:NatGd} and their radiative thermal neutron capture 
             cross-sections~\cite{Mughabghab2006:NeutRes}.}
\centering
    \begin{tabular}{rcc}
        \hline
        Isotope & Abundance[\%] & Cross-section[b]\\
%                 & ${}$ [\%] &  [b]\\
        \hline
        ${}^{152}$Gd & 0.200 & 735    \\
        ${}^{154}$Gd & 2.18  & 85   \\
        {\bf ${}^{155}$Gd} & {\bf 14.80} & {\bf 60900}  \\
        ${}^{156}$Gd & 20.47 & 1.8   \\
        {\bf ${}^{157}$Gd} & {\bf 15.65} & {\bf 254000} \\
        ${}^{158}$Gd & 24.84 & 2.2   \\
        ${}^{160}$Gd & 21.86 & 1.4   \\  
%           Others & $\sim$69.55 &    \\  
        \hline
    \end{tabular}
    \label{tab:gdisotopes}
    \end{table}

The most frequent isotopes,   $^{\rm 155}$Gd and $^{\rm 157}$Gd, are as well featuring the
 large cross section of thermal neutron capture. 
Therefore, the required MC model for $^{\rm nat}$Gd requires the modelling of the $\gamma$-ray emission from not only $^{\rm 157}$Gd  \cite{Hagiwara:2018kmr} but also $^{\rm 155}$Gd. 
  
We measured the $\gamma$-ray energy spectrum from the radiative
 thermal neutron capture on an  enriched $^{\rm 155}$Gd sample and a $^{\rm nat}$Gd film with the 
germanium (Ge) spectrometer of the Accurate Neutron-Nucleus Reaction Measurement Instrument 
(ANNRI)~\cite{Igashira2009:MLF-BL04, Kin2011:ANNRI, Kino2011:ANNRI, Kimura2012:ANNRI, 
Kino2014:ANNRI}. 
%The high-purity Ge-detectors of the ANNRI \cite{Hagiwara:2018kmr}, located at Beam Line No. 4~\cite{Igashira2009:MLF-BL04} of the MLF 
%provide for measuring neutron-nucleus interactions with its fine energy resolution. 
The incident pulsed neutron beam  from the Japan 
Spallation Neutron Source (JSNS) at the Material and Life Science Experimental Facility (MLF) of the Japan 
Proton Accelerator Research Complex (J-PARC)~\cite{Nagamiya2012:JPARC} and the good 
$\gamma$-ray energy resolution, high statistics and low background makes ANNRI a favorable spectrometer for our intended study ~\cite{Hagiwara:2018kmr, Igashira2009:MLF-BL04}.

%%  2019.0822 Sakuda 1 copied the following paragraph from Gd157 paper, whose texts  must be modified. 
%There have been several publications on measured $\gamma$-ray spectra from Gd($n,\gamma$) 
%reactions for neutron energies ranging from meV to MeV~\cite{Groshev1962:GdGam, 
%Bollinger1970:NeutARC, Voignier1986:GamSpecFromNeutInt, Ali1994:Gd157, Valenta2015:GdTwoStepGamCasc}. 
The Detector for Advanced Neutron Capture Experiments (DANCE) at the Los Alamos Neutron 
Science Center (LANSCE) has extensively studied the $\gamma$-ray energy spectra from the radiative 
neutron capture reaction at various multiplicities in the neutron kinetic
energy range from 1 to 300 eV for both $^{155}$Gd and $^{157}$Gd 
targets~\cite{Baramsai2013:DANCE, Chyzh:DANCE, Kroll2013:DANCE}. They compared their $\gamma$-ray spectra
to MC simulations with the DICEBOX package~\cite{Becvar1998:DICEBOX} and showed fair agreement. 
Concerning the measurements in the thermal energy region, Groshev \textit{et al.}~\cite{Groshev1959:GdGam} measured
prompt  $\gamma$ rays from the neutron capture on $^{155}$Gd and $^{157}$Gd and tabulated the $\gamma$-ray energy, intensity values and decay schemes in great details. Valenta \textit{et al.}~\cite{Valenta2015:GdTwoStepGamCasc} measured 
the two-step cascade (TSC) $\gamma$ rays, following the thermal neutron capture on $^{155}$Gd and $^{157}$Gd,  by a pair of HPGe detectors and studied the effect of  the M1 or  E2  transitions in addition to the  E1 transitions in the TSC spectra. 

 We performed a series of measurements of the prompt $\gamma$ rays covering 
almost the full spectrum from 0.11 MeV to 9 MeV  from the capture reaction on $^{155, 157}$Gd and $^{nat}$Gd
at thermal neutron energies.  As we demonstrated in Fig. 12 of the previous publication~\cite{Hagiwara:2018kmr} and also Fig.7 of this report, it is very important to measure  the full  $\gamma$-ray spectrum from the capture  in order to study the photon strength function and the nuclear level density, which are the important properties of the Gd nucleus~\footnote{The high-energy part  of the $\gamma$-ray spectrum above 4 MeV is dominated by the first $\gamma$-ray transition from the resonance and is sensitive to the shape of the E1 photon strength function; the low-energy part of the spectrum below 4 MeV is mainly contributed to by 
the subsequent cascade $\gamma$ rays.}.  
  Based on our data and a Geant4-based detector simulation~\cite{Agostinelli2003:Geant4, Allison2006:Geant4} of our setup, we 
 developed a Monte Carlo (MC) model to generate the full $\gamma$-ray spectrum from the thermal 
${}^{157, 155, nat}$Gd($n$,$\gamma$) reaction.  The $\gamma$-ray spectrum and its corresponding MC model (ANNRI-Gd model) for $^{\rm 157}$Gd has already been discussed in Ref.~\cite{Hagiwara:2018kmr}. 
 
 %%Sakuda 191122 Daya Bay NIMA940
%There are several models and measurements to describe the \00 energy spectra, such as the model in Geant4, the model based on the Nuclear Data Sheet, and measurements by several groups. %Large discrepancies are found among these models and measurements.
 %% Sakuda 1 end ---------------------------

 In this report, we present the $\gamma$-ray energy spectra from  the $^{\rm 155}$Gd(n, $\gamma$) and $^{\rm nat}$Gd(n, $\gamma$) reactions, modify our ANNRI-Gd model with the contribution from $^{\rm 155}$Gd and present our final MC performance for $^{\rm nat}$Gd(n, $\gamma$) to be used by any neutrino or other experiments involving the measuriement of  $\gamma$-ray signals from the thermal neutron capture on Gd. 
%  However, the neutrino experiments use natural Gd ($^{\rm nat}$Gd), which comprise of 14.80\% of $^{\rm 155}$Gd, and 15.65\% of $^{\rm 157}$Gd. 

%  The consistency of the results from the devised model is checked among all the 14 germanium crystals, at the level of 15\%  spectral shape deviation at 0.2 MeV binning.

% We report the results from the spectrometry of ${}^{155}$Gd and natural Gd which comprise of 

%==================================================================================================%
\section{Experiment and Data Analysis}
\label{sec:exp}
%==================================================================================================%

The 300 kW beam of 3 GeV protons from the JSNS facility in double-bunch 
mode and a frequency of 25 Hz was incident on a primary target of mercury, 
producing neutrons. The neutron beam thus produced consist of neutron pulses in double 
bunch mode, each 100 ns wide, with  600 ns spacing every 40 ms.
% a double of 100 ns wide neutron beam bunches with 600 ns spacing every 40 ms. 
The ANNRI spectrometer is located 21.5 m away from the neutron beam 
source. It comprises two germanium cluster detectors with anti-coincidence shields 
made of bismuth germanium oxide (BGO) and eight co-axial germanium detectors. The target for
neutron capture is positioned in line with the beam, at 13.4 cm from each of the two cluster detectors
on its either side along the vertical plane.

% midway between the two vertically clusters,  
In this report, we used only data taken with the cluster detectors which cover  15\% of solid angle.
%We collected data during the absence of seven of the co-axial detectors, thus covering 15\% of solid angle. 
Each cluster consists of seven Ge-crystals in a hexagonal 
arrangement, details of which can be found in Ref.~\cite{Hagiwara:2018kmr}.

%__________________________________________________________________________________________________%

\begin{figure}[h]
  \centering
  \begin{minipage}{0.48\textwidth}
  \includegraphics[trim=0.1cm 0.1cm 1.0cm 0.1cm, clip=true, 
  width=\textwidth]{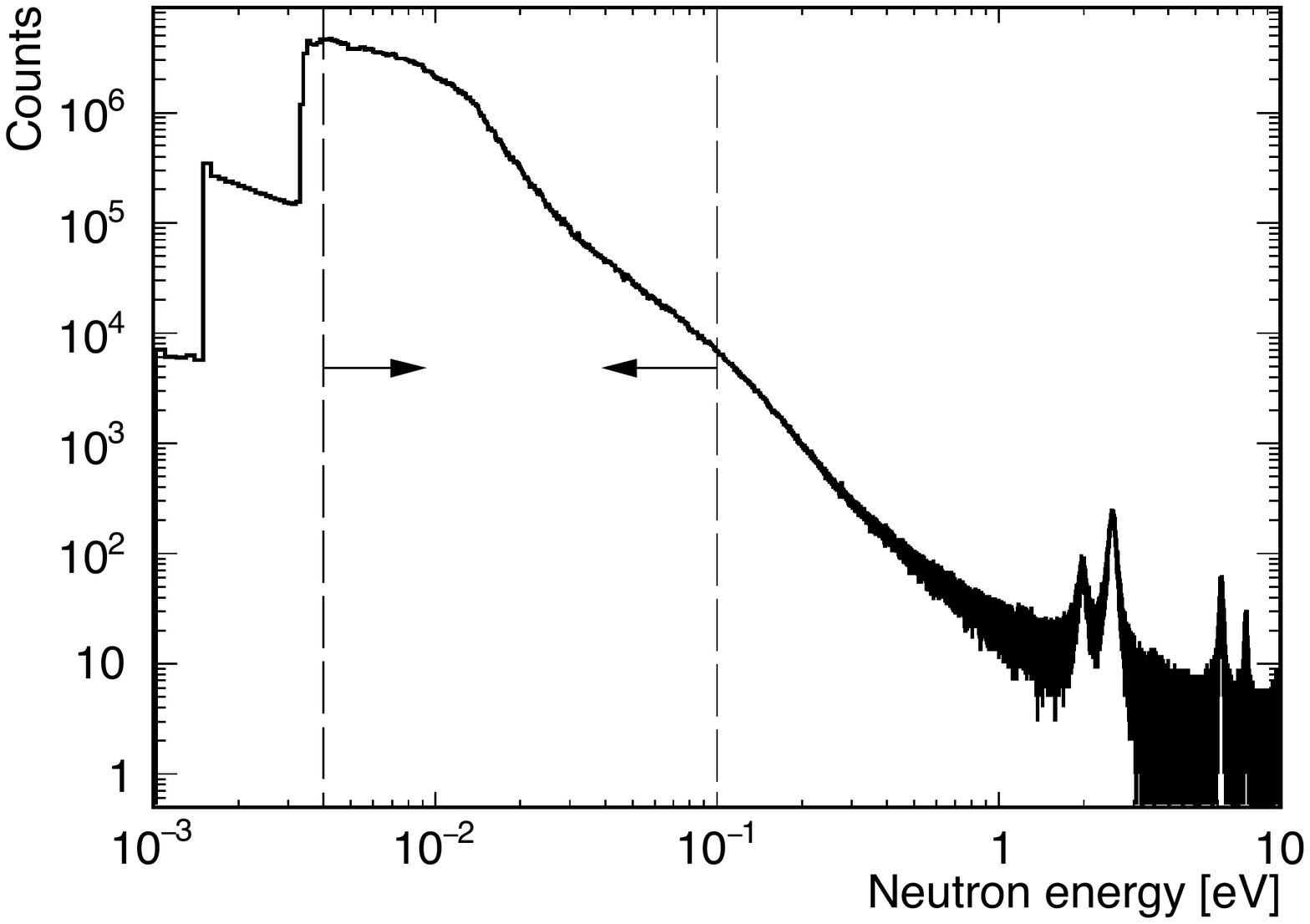}
  \end{minipage}%
  ~
  \begin{minipage}{0.48\textwidth}
  \includegraphics[trim=0.1cm 0.1cm 1.0cm 0.1cm, clip=true, 
  width=\textwidth]{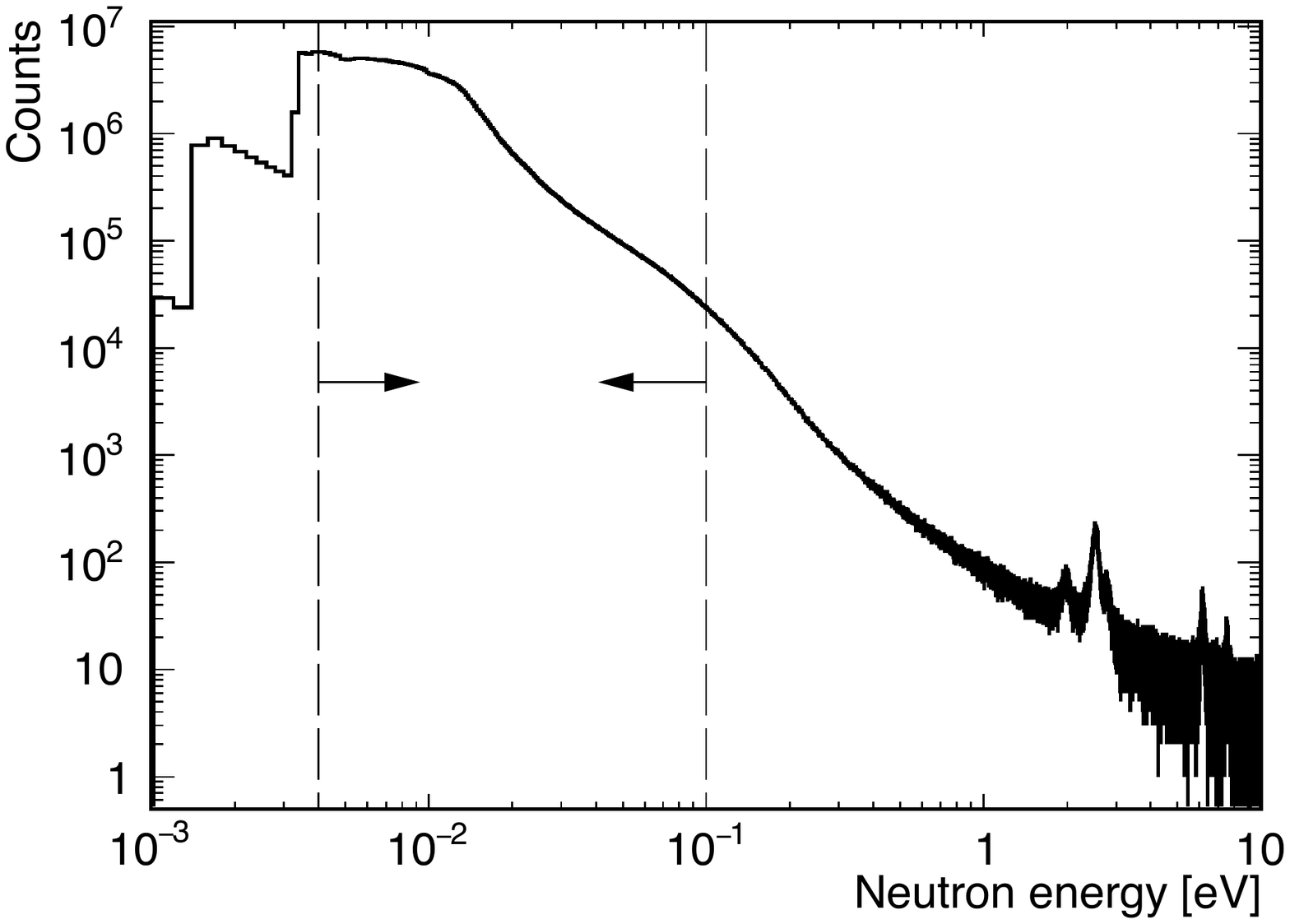}
  \end{minipage}%
   \caption{Energy spectrum of neutrons as obtained with the observed 
             neutron time-of-flight according to Eq.~(\ref{eq:NeutronToFtoErgy}) for the ${}^{155}$Gd target (left) and the ${}^{nat}$Gd target (right).}
    \label{fig:NSpec}  
\end{figure}

From the neutron time-of-flight $T_{\mathrm{TOF}}$ recorded for each event 
we calculated the neutron kinetic energy $E_n$ as
\begin{equation}
  E_n = m_n(L/T_\mathrm{TOF})^2 / 2 \, ,
  \label{eq:NeutronToFtoErgy}
\end{equation}
where $m_n$ is the neutron mass and $L$ is the 21.5 m distance 
between neutron source and target. The resulting neutron energy spectra are shown in 
Fig.~\ref{fig:NSpec}. Since we study the $\gamma$-ray 
spectrum solely from thermal neutron capture on ${}^{155}$Gd and ${}^{nat}$Gd, we only selected events from neutrons 
in the kinetic energy range $[4, 100]$ meV for the present analysis.

%__________________________________________________________________________________________________%
% FIGURE BEGIN

% FIGURE END
%___________

%%Sakuda 191104  %%The time-of-flight (TOF) information enables a precise selection of  neutron events in the energy range from 4 to 100 meV  for the analysis. 
The obtained data covers the energy region of $\gamma$ rays from 0.11 MeV to about 9 MeV with observed 
 $\gamma$-ray multiplicities ($M$) one to three. 
The energies of the emitted $\gamma$ rays are recorded by each of the crystals. A threshold of 100 keV
 is set for each of the cluster detectors. 
 For the event classification, we assign a multiplicity value $M$ and a hit value $H$  to each recorded event. We defined the multiplicity 
$M$  as the combined number of isolated sub-clusters of hit Ge crystals at the upper 
and the lower  clusters. A sub-cluster is formed by the neighboring hit Ge crystals 
and can be of size $\geq 1$. The hit value $H$ describes the total number of Ge crystals hit 
in the event. The multiplicity $M$ represents the number of observed $\gamma$ rays, while the hit value $H$ is a measure of
the lateral spread of  $\gamma$ rays. The details of the event class are described in Ref.~\cite{Hagiwara:2018kmr}. 
%We follow the same algorithm as described in paper \cite{Hagiwara:2018kmr} to group the adjacent or neighbouring crystals receiving the $%%\gamma$ rays (hit crystals) into a subcluster.  This helps to avoid considering of multiple $\gamma$ rays, when a single one deposits its energy in %more than one crystal. So, the value of hits (H) basically gives the sum of the number of crystals considered in a subcluster, while our %%%\emph{detectable, obesrved?} multiplicity (M) is given by the number of sub-clusters.  
% \textcolor{red}
The fraction of the data collected in each event class are reflected in the barcharts in Fig.~A1.

We used radioactive sources ($^{60}$Co, $^{137}$Cs, and $^{152}$Eu) and $^{35}$Cl($n$,$\gamma$) to calibrate the detector, 
and determine the detection efficiency of the spectrometer for 
$\gamma$-rays at energies from 0.3 to 8.5 MeV, as decribed in details in Ref.~\cite{Hagiwara:2018kmr}.

% \begin{table}[H]
% \centering
% % \scriptsize
% \begin{tabular}{|c|c||c|c|}
%      
%       \hline 
%       \multicolumn{2}{|c||}{Class} & \multicolumn{2}{c|}{Data Fraction [\%]}\\
%       \hline
%       M & H &  $^{155}$Gd & $^{Nat}$Gd\\
%       \hline \hline
%       \multirow{4}{*}{1}& 1 & ? & ? \\
%      \cline{2-4}
% 			& 2 &   &  \\
%      \cline{2-4}
% 			& 3 &   &  \\
%      \cline{2-4}
% 			& 4 &   &  \\
%       \hline
% 
% % 	\multicolumn{2}{|c||}{Sum } & ??$\times10^??$ & ??$\times10^??$\\
% %       \hline  \hline
%       \multirow{3}{*}{2}& 2 &   &  \\
%      \cline{2-4}
% 			& 3 &   &  \\
%      \cline{2-4}
% 			& 4 &   &  \\
%       \hline
%       \multirow{2}{*}{3}& 3 &   &  \\
%      \cline{2-4}
% 			& 4 &   &  \\
%       \hline     \hline
% 	\multicolumn{2}{|c||}{Total events } & ??$\times10^??$ & ??$\times10^??$\\
%       \hline
% 
% \end{tabular}
% \label{tab:mxhxtab}
% \end{table}

We measured the thermal neutron capture on a gadolinium (Gd$_2$O$_3$) target enriched with
${}^{155}$Gd (91.85\%) in December 2014 and the natural Gd (99.9\% pure metal film) in March 2013. 
The weights of the targets, i.e., $^{\rm 155}$Gd and $^{\rm 157}$Gd powder were 26.4 mg and 28.9 mg respectively, spread across an area of 1cm x 1cm in a teflon envelope. The film of the natural gadolinium target was 5mm x 5mm x 10 $\mu$m (and 20 $\mu$m) in dimensions. 
The isotopic composition of our enriched gadolinium sample is given in Table~\ref{tab:gdabundance}. 

\begin{table}[H]
\caption{Isotopic compositions of the Gd$_2$O$_3$ targets.}
\centering
    \begin{tabular}{|r|c|c|c|c|c|c|c|}
        \hline
        Isotope: & $^{152}$Gd & $^{154}$Gd &$^{155}$Gd & $^{156}$Gd &$^{157}$Gd & $^{158}$Gd &$^{160}$Gd\\
%                 & ${}$ [\%] &  [b]\\
        
%         (Natural Abundance) & (0.2\%) & (2.18\%) & (14.8\%) & (20.5\%) & (15.7\%) &(24.8\%) & (21.9\%)    \\
        \hline ${}^{155}$Gd$_2$O$_3$ &$<$0.02  & 0.5 & 91.9($\pm$0.3) & 5.87 & 0.81 &0.65 &0.27  \\
        \hline ${}^{157}$Gd$_2$O$_3$ &$<$0.01  & 0.05 & 0.3 & 1.63 & 88.4($\pm$0.2) &9.02 &0.6  \\  
        \hline
    \end{tabular}
    \label{tab:gdabundance}
    \end{table}

\begin{figure}[H]
  \begin{center}
\includegraphics[width=1.\textwidth]{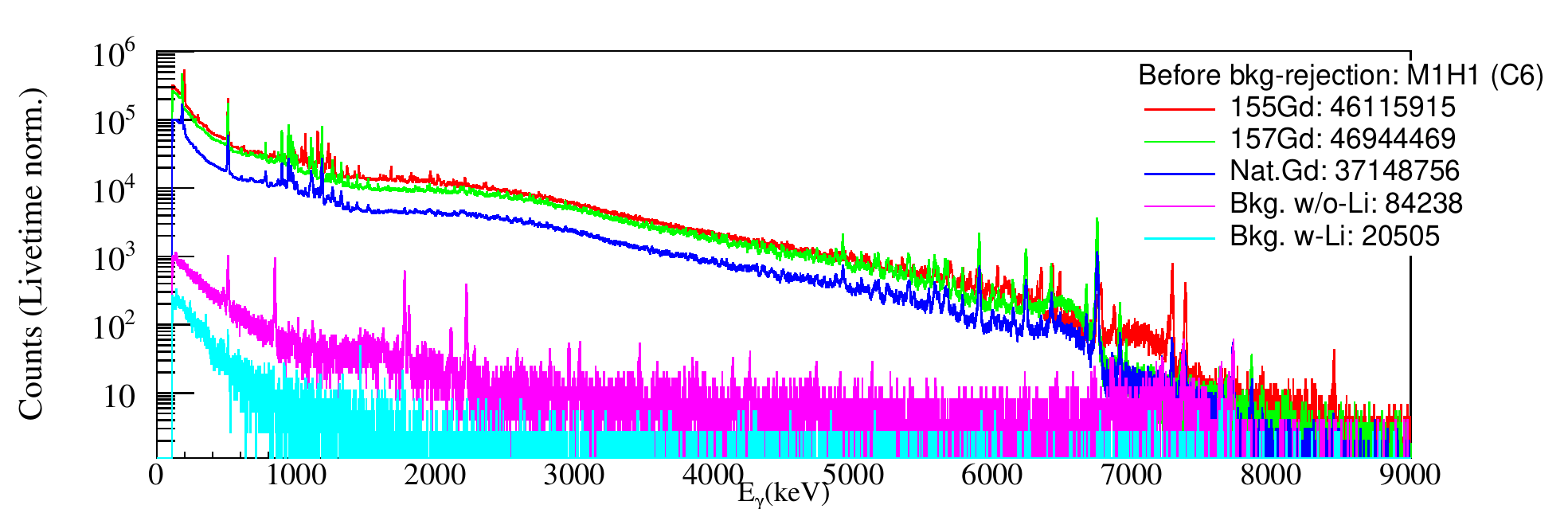}
    \caption{Energy spectra for M1H1 (1$\gamma$ and 1 hit) events obtained with neutron beam on the targets $^{155}$Gd, $^{157}$Gd and natural gadolinium, and the blank target holder  as recorded in 2013 (w/o LiF) and 2014 (with LiF). The numbers show the data statistics in each case.}
    \label{fig:b4bkg}
  \end{center}
\end{figure}

\begin{figure}[H]
  \begin{center}
   \includegraphics[width=1.\textwidth]{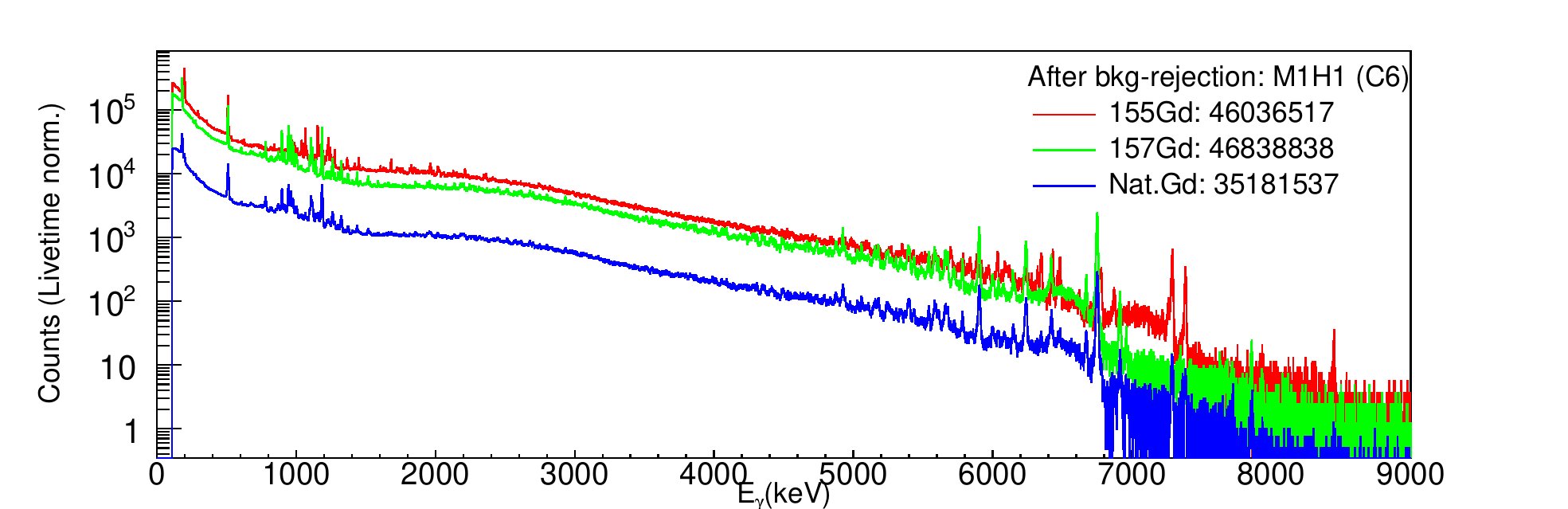}
    \caption{Energy spectra for M1H1 events obtained with neutron beam on the targets $^{155}$Gd, $^{157}$Gd and natural gadolinium, after subtracting the background. The numbers show the data statistics in each case.}
    \label{fig:aftrbkg}
  \end{center}
\end{figure}

In 2014, the beam pipe included an additional layer of LiF ($\sim$1 cm thickness) was included to the beam pipe 
to reduce the $\gamma$ rays from neutron capture on the aluminium of the beam pipe. Therefore, the data-taking with ${}^{\rm nat}$Gd was subject to more background events (without the LiF layer) than that of ${}^{155, 157}$Gd. 
The background $\gamma$-ray energy spectra which were observed by one of the crystals (C6) for M1H1 events (1$\gamma$ and 1 hit)
with the empty target holder  at two different periods  in the neutron beam are shown in Fig.~\ref{fig:b4bkg}. 
The  $\gamma$-ray energy spectra for M1H1 events with the three target materials, $^{155}$Gd, $^{157}$Gd (2014) and $^{\rm nat}$Gd (2013)  are also shown in Fig.~\ref{fig:b4bkg}. The histograms shown are normalized with reference to the live time of $^{155}$Gd data set. The differences in the observed count rates are due to the differences in the target masses  ($\times$ cross section) used for the three measurements. 
%Sakuda adds one line which describes the size of the background.
The size of the background is less than 0.1\% for the data of  the $^{155}$Gd target and less than 1\% for those of the $^{\rm nat}$Gd target. 
The background is accordingly subtracted for each data set and the resulting energy spectra for the three targets are shown in Fig.~\ref{fig:aftrbkg}.

%_______________________________________________________________________________________%

% (comment) We may give the parameters of the target. This will clarify the difference in the statistics.  If you think it also good, then please correct English: The weight of each 155Gd and 157Gd powder was 26.4 mg and 28.9 mg spread over 1cm x 1cm teflon sheet. The size of a natural gadolinium target was 5mm x 5mm x 10 micron (and 20 microns).  

%__________________________________________________________________________________________________%
% FIGURE BEGIN
% FIGURE END
%__________________________________________________________________________________________________%

%__________________________________________________________________________________________________%
% FIGURE BEGIN
% FIGURE END
%__________________________________________________________________________________________________%

The $\gamma$-ray energy spectrum from neutron capture on natural gadolinium is dominated by that from its two main isotopes, $^{155}$Gd and $^{157}$Gd, with fractions of 18.5\% and 81.5\%, respectively. 
The contributions of other isotopes are negligible. 

%%Sentence below is wrong. 2019 April 15. -M.Sakuda

%%Their content 
% The fraction of $^{155}$Gd and $^{157}$Gd in natural gadolinium 
%%adds up to only $\sim$30\%  in natural gadolinium. This explains the  lower statistics for the natural Gd spectrum normalized over the same %%%livetime\footnote{Details of the corrected livetime calculation after subtraction of the dead time can be found here in~\cite{Hagiwara:2019PTEP}}, %%than that of the  $^{155}$Gd and the $^{157}$Gd enriched targets, in the fig.~\ref{fig:b4bkg} and \ref{fig:aftrbkg}. 

% {\bf Compare Combined data}

\begin{figure}[H]
  \begin{center}
   \includegraphics[width=0.8\textwidth]{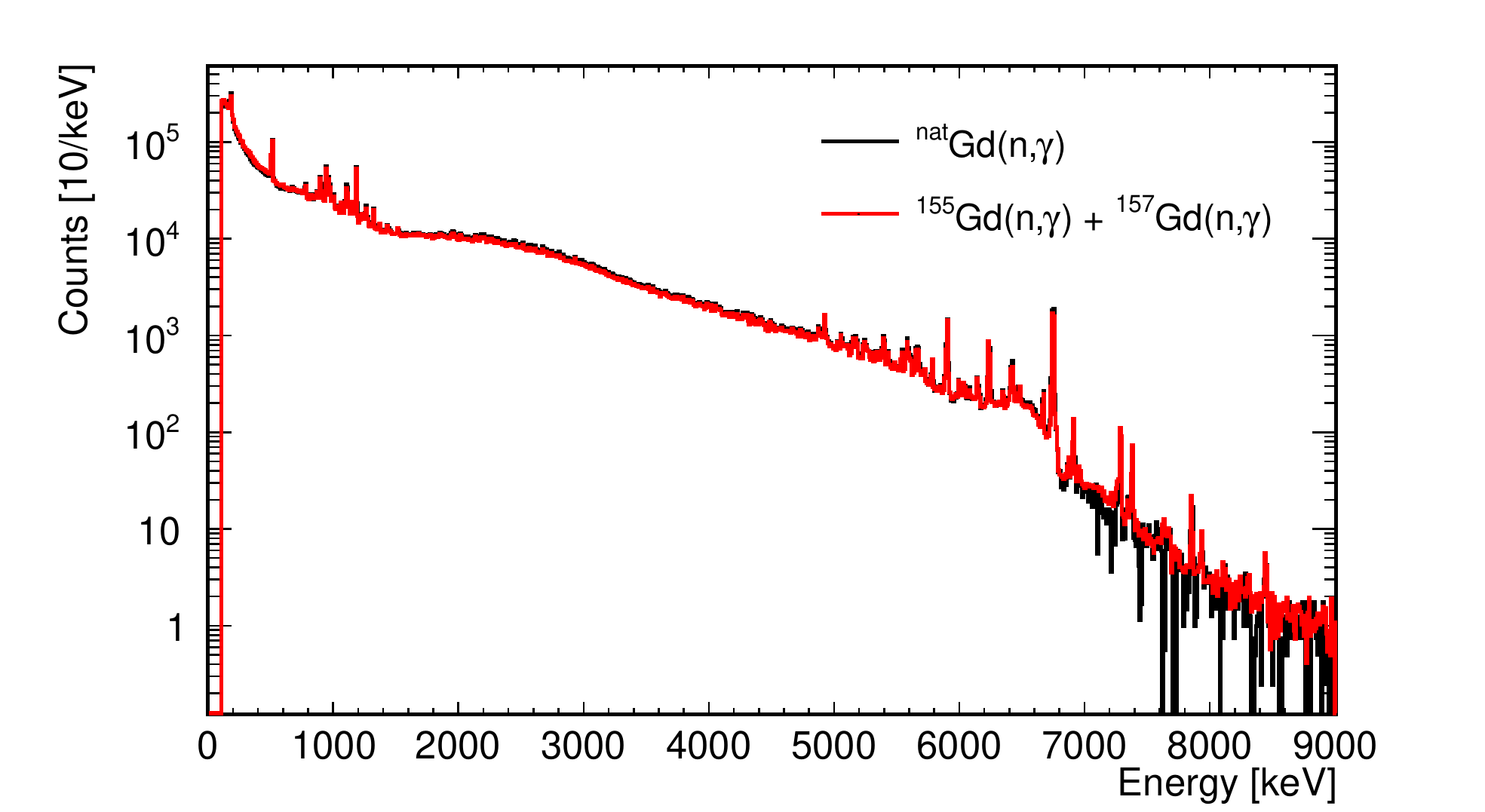}
    \caption{Comparison of the combined energy spectra of $^{155}$Gd and $^{157}$Gd (red) with that of the natural gadolinium (black). }
    \label{fig:datacomp}
  \end{center}
\end{figure}

% As already said earlier, the gamma emission by natural gadolinium on neutron capture is dominated by the same as for $^{155}$Gd and $^{157}$Gd. 
The spectra taken separately for the pure $^{155}$Gd and $^{157}$Gd samples must be consistent with that of the $^{nat}$Gd film, when they are combined in the corresponding proportions. This was checked and confirmed in Fig.~\ref{fig:datacomp}, where excellent agreement is found between the two spectra  (red and black). 
%   Single energy hit spectrum or M1H1 most dominant: $\sim$70\%

%__________________________________________________________________________________________________%
% FIGURE BEGIN
% FIGURE END
%__________________________________________________________________________________________________%
 
 \section{Update of ANNRI-Gd Model}
The MC model for $^{\rm 157}$Gd has been already described in Ref.~\cite{Hagiwara:2018kmr}. We now develop a MC model for $^{\rm 155}$Gd, following the same approach of a separate treatment of the discrete and the continuum part of the spectrum~\cite{GLG4sim2005:HP, Hagiwara:2018kmr}. 

\begin{figure}[H]
  \begin{center}
    \includegraphics[trim=0.1cm 0.1cm 0.5cm 0.5cm, clip=true,width=0.48\textwidth] 
    {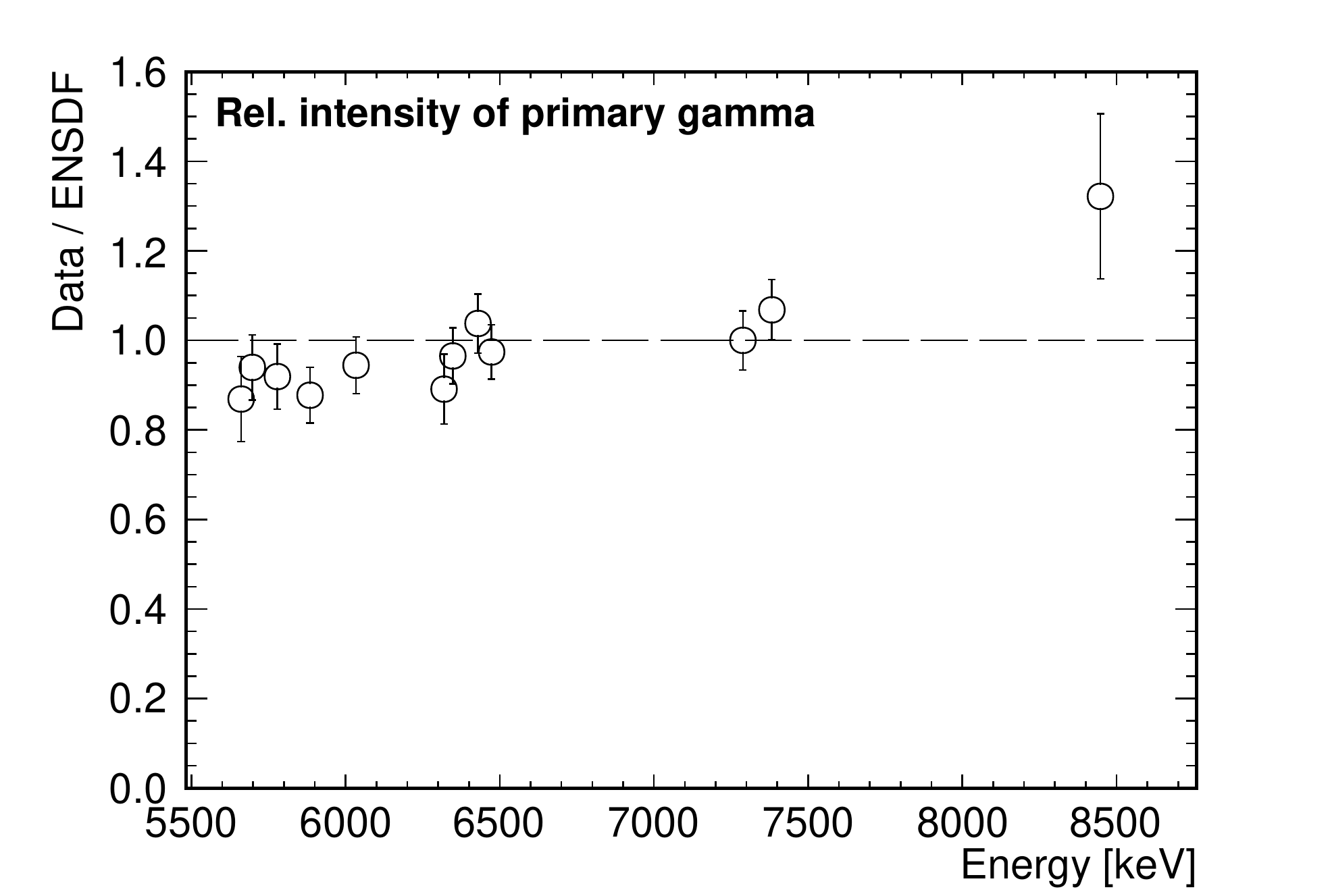}
    \includegraphics[trim=0.1cm 0.1cm 0.5cm 0.5cm, clip=true,width=0.48\textwidth] 
    {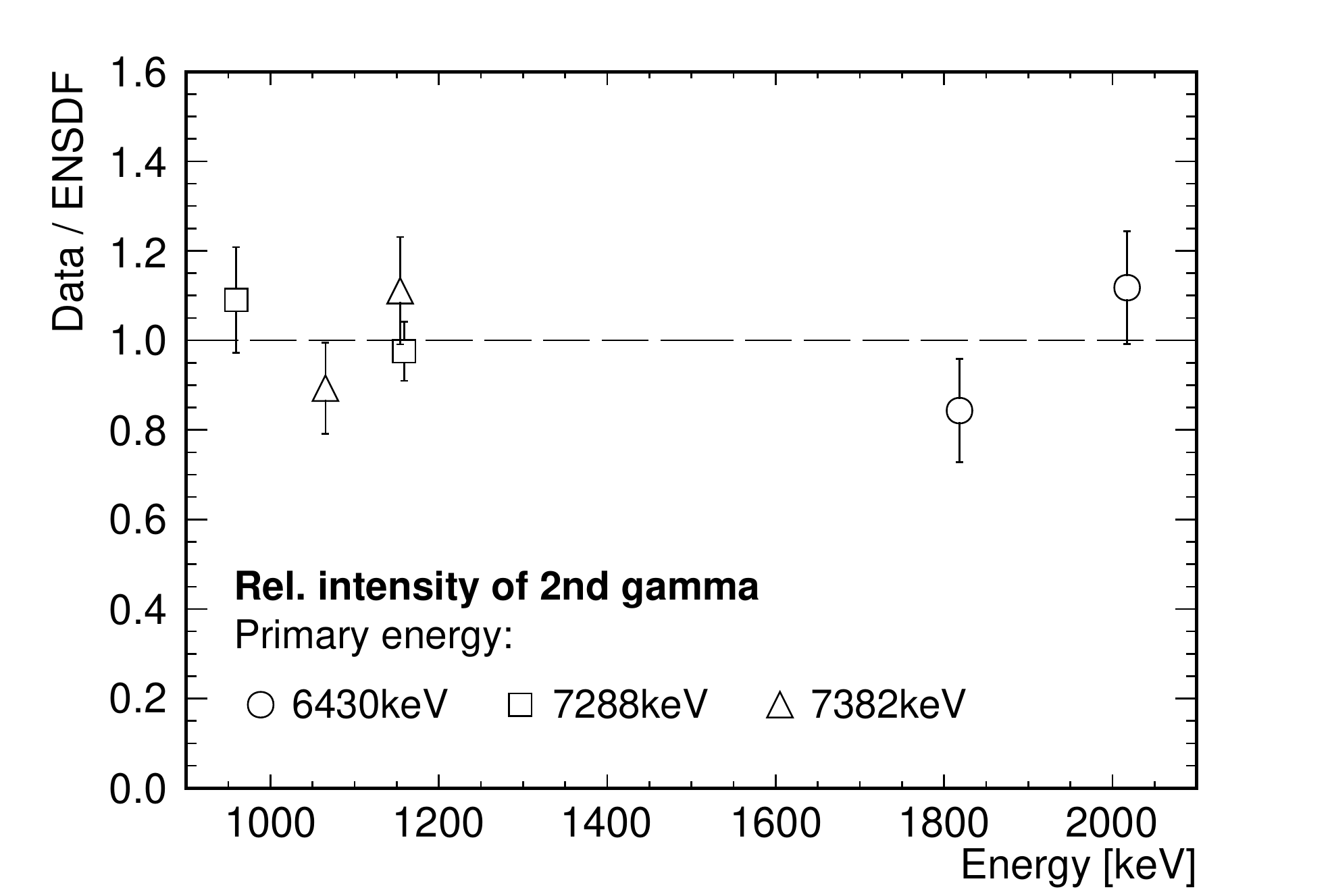}
    \caption{Relative intensity of the primary peaks (left) and the secondary $\gamma$ rays (right) compared with the values from Nuclear Data Sheets for $A=156$~\cite{Reich:2013bqa}.}
    \label{fig:relint}
  \end{center}
\end{figure}

For the thermal neutron capture on  $^{\rm 155}$Gd in an s-wave, the resonance state is 8.536 MeV ($J^\pi = 2^-$) of  $^{\rm 156}$Gd.  
The resonance energy for the neutron is 26.8$\pm$0.2  meV and the radiative width is 108$\pm$1 meV~\cite{Mughabghab2006:NeutRes}.
We identified and measured the photo peak intensities of 12 discrete $\gamma$ rays for $^{\rm 155}$Gd(n, $\gamma$) above 5~MeV as listed in Table~\ref{tab:discPeaks}. The single and double escape peaks were excluded before analysing these peaks.  The direct transition of the resonance state  
($J^\pi = 2^-$) to the ground state ($J^\pi = 0^+$) is largely suppressed   compared to the transition from 8.536 MeV ($J^\pi = 2^-$)   to 0.089 MeV ($J^\pi = 2^+$), emitting a 8.448-MeV $\gamma$ ray. The tabulated values of the energies are taken from Ref.~\cite{Reich:2013bqa}. In case of overlapping peaks in our data spectrum, we mention the means of the primary $\gamma$-ray energies with their combined intensities. The discrete $\gamma$-ray emissions above 5 MeV are expected to arise mostly from the first transition and are hence referred to as `primary' $\gamma$ rays. By tagging the events with each of these primary $\gamma$ rays, we obtained the intensities of the secondary $\gamma$ rays. We found them in fair agreement with the values from Nuclear Data Sheets for $A=156$~\cite{Reich:2013bqa}, as displayed in Fig.~\ref{fig:relint}. Details of the comparison methods are described in  Ref.~\cite{Hagiwara:2018kmr}.
The relative intensities of these discrete peaks add up to 2.78$\pm$0.02\% of the data spectrum. 
% The rest but dominant contribution of 97.22$\pm$0.02\% comes from the continuum region of the energy levels in $^{\rm 156}$Gd. 

% \cite{Nica2017:NuclDataGd158}

%__________________________________________________________________________________________________%
% TABLE BEGIN
\begin{table}[t!b]
  \centering
  \caption{List of the 12 discrete peaks from primary $\gamma$ rays we identified in our data. The 
          stated energies are taken from Ref.~\cite{Reich:2013bqa}, rounded to nearest keV. In four 
          cases the table lists the unweighted mean energy of known peaks that overlap in 
          our data: 
           (i)   6474 keV combining 6482 keV and 6466 keV, 
           (ii)  6348 keV combining 6349 keV and 6345 keV, 
           (iii) 5885 keV combining 5889 keV and 5884 keV, as 
                well as
           (iv)  5779 keV combining 5774 keV and 5786 keV.}
  \label{tab:discPeaks}
  \begin{tabular}{|r|c|c|c|c|}
    \hline
    \multicolumn{4}{|c|}{$\gamma$-ray energy [keV]} & Intensity \% \\
    \cline{1-4}
    &Primary & \multicolumn{2}{|c|}{Secondary} & [$10^{-2}$] \\ 
    \hline 
		     1  &  8448 	       & --    & --  &  1.8   $\pm$ 0.2  \\ \hline
     \multirow{2}{*}{2}	&\multirow{2}{*}{7382} & 1154  &  -- &  12.7  $\pm$ 1.4  \\ \cline{3-5}
			&                      & 1065  &  -- &  10.6  $\pm$ 1.2  \\ \hline
     \multirow{2}{*}{3}	&\multirow{2}{*}{7288} & 1158  &  -- &  34.8  $\pm$ 2.4  \\ \cline{3-5}
			&                      & 959   &  199&  10.5  $\pm$ 1.1  \\ \hline
 		     4  &  6474                & 1964  &  -- &  35.2  $\pm$ 0.7  \\ \hline
     \multirow{2}{*}{5}	&\multirow{2}{*}{6430} & 2017  &  -- &  20.7  $\pm$ 2.2  \\ \cline{3-5}
			&                      & 1818  &  199&  11.7  $\pm$ 1.5  \\ \hline
     \multirow{4}{*}{6}	&\multirow{4}{*}{6348} & 2188  &  -- &  12.1  $\pm$ 1.7  \\ \cline{3-5}
			&                      & 2097  &  199&  9.8   $\pm$ 1.6  \\ \cline{3-5}
			&     & \multirow{2}{*}{1036}  &  1154&  4.6  $\pm$ 0.8  \\ \cline{4-5} 
			&                      &	& 1065&  3.8  $\pm$ 0.7  \\ \hline
 		     7  &  6319                & 2127  &  -- &  9.4   $\pm$ 0.5  \\ \hline
     \multirow{2}{*}{8}	&\multirow{2}{*}{6034} & 2412  &  -- &  14.0  $\pm$ 1.7  \\ \cline{3-5}
			&                      & 2213  &  199&  6.4   $\pm$ 1.0  \\ \hline
     \multirow{2}{*}{9}	&\multirow{2}{*}{5885} & 2563  &  -- &  9.0   $\pm$ 2.1  \\ \cline{3-5}
			&                      & 2364  &  199&  8.4   $\pm$ 2.1  \\ \hline
 		     10 &  5779                & 2672  &  -- &  18.8  $\pm$ 0.8  \\ \hline
 		     11 &  5698                & 2749  &  -- &  28.6  $\pm$ 0.8  \\ \hline
 		     12 &  5661                & 2786  &  -- &  15.4  $\pm$ 0.7  \\ \hline
 \end{tabular}
\end{table}

%__________________________________________________________________________________________________%
% FIGURE BEGIN
% FIGURE END
%__________________________________________________________________________________________________%

 For the modelling of the continuum part, we  compute the probability $P(E_a,E_b)$ for E1 transitions with $E_\gamma = E_a - E_b$  in terms of  transmission coefficient $T_{E1}(E_\gamma)$ and the number of levels $\rho$(E$_b$)$\delta$E$_b$ as

%\begin{equation}

%P_a(E_a,E_b)\Delta E_b =\frac{T(E_a,E_b)[\rho(E_b)\Delta E_b]}{\int_0^{E_a} T(E_a,E_b) \rho(E_b) dE_b}\  

% \end{equation}

\begin{figure}[h]
  \begin{center}
    \includegraphics[trim=0.1cm 0.1cm 0.5cm 0.5cm, clip=true,width=0.48\textwidth] 
    {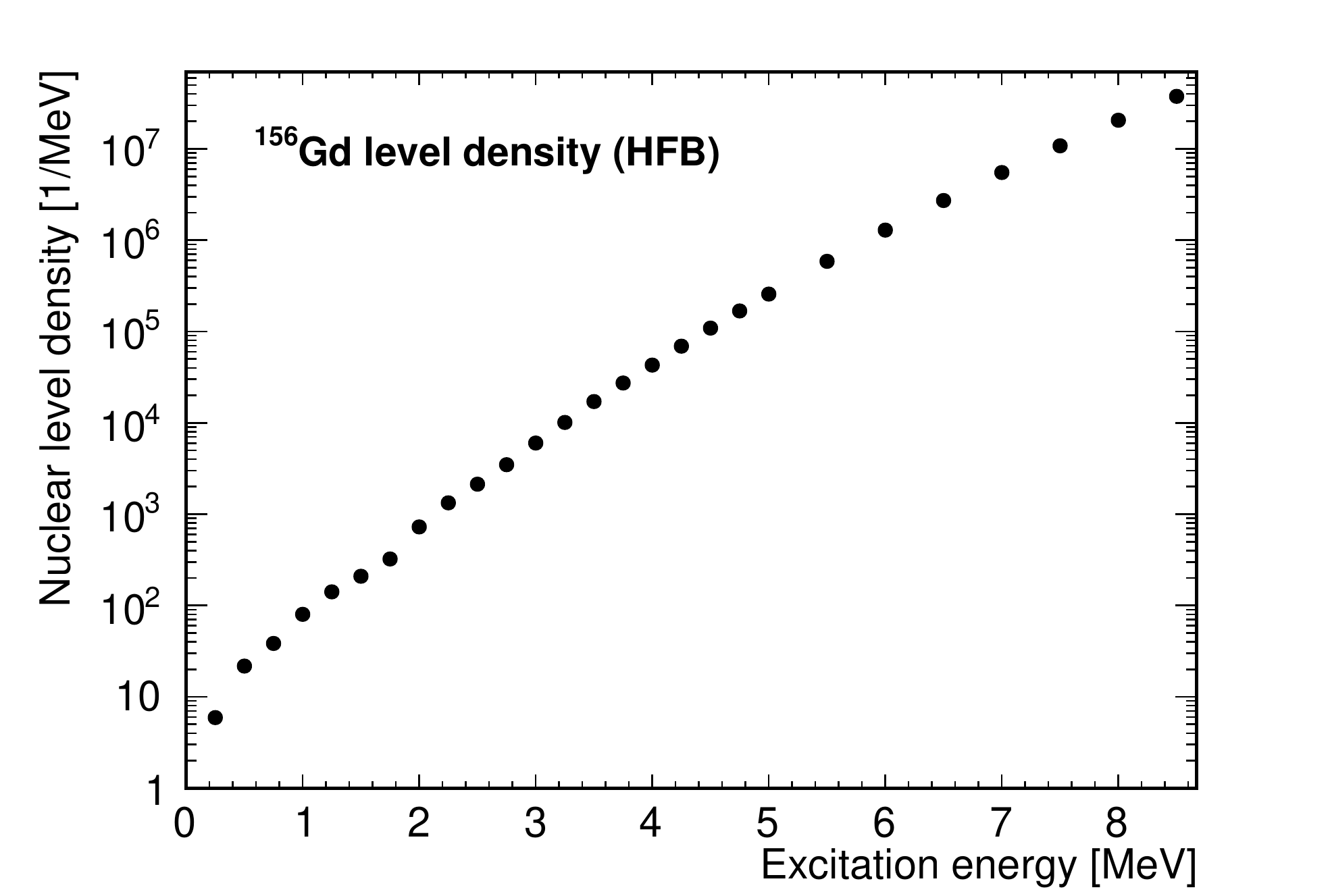}
    \includegraphics[trim=0.1cm 0.1cm 0.5cm 0.5cm, clip=true,width=0.48\textwidth] 
    {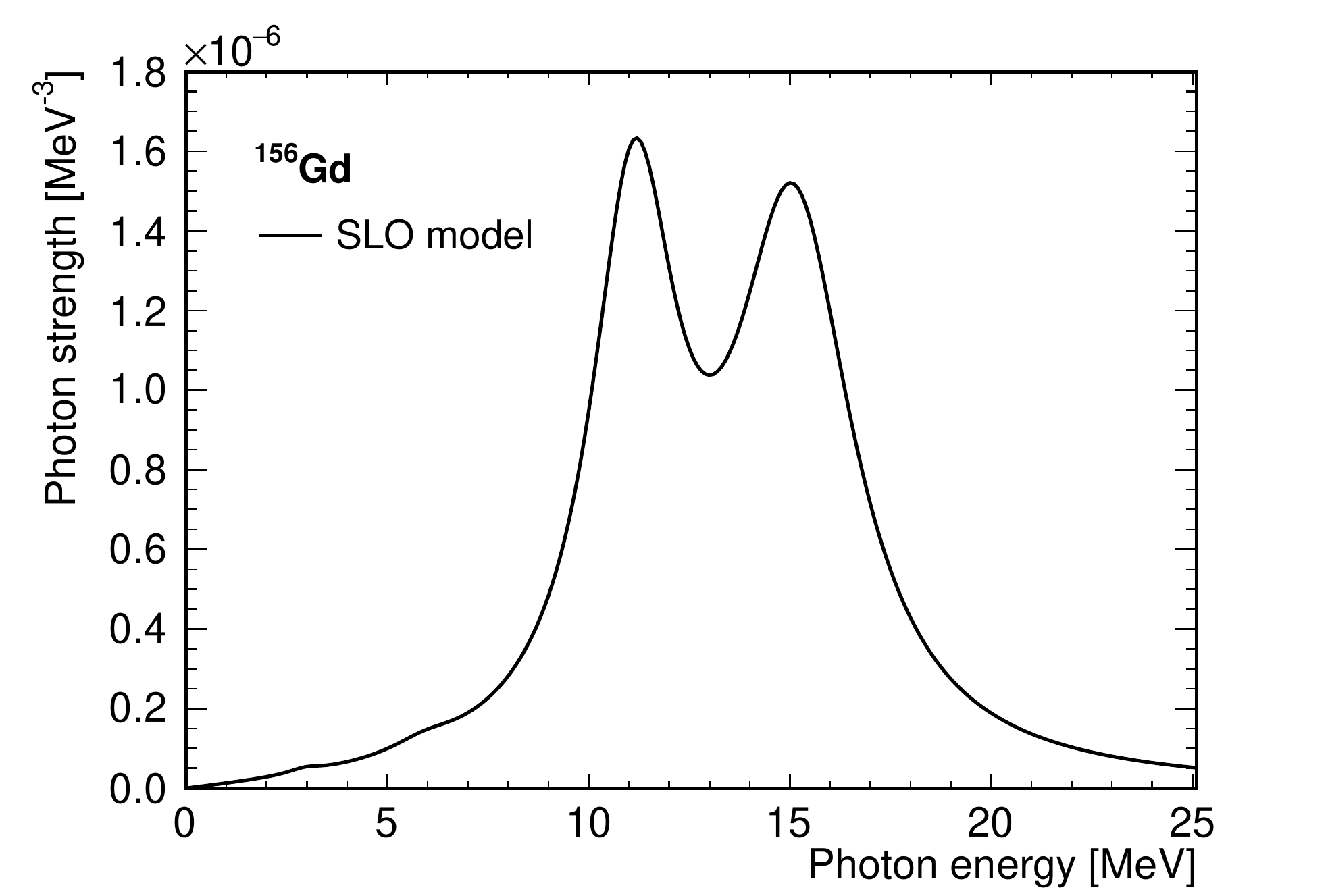}
%     {PSF_2parameter_156Gd.pdf}
    \caption{Left: Tabulated values~\cite{Capote2009:RIPL3} for the NLD of ${}^{156}$Gd from 
           computations based on the HFB method~\cite{Goriely2007:HFB, Goriely2008:HFB}. Right: The E1 PSFs for ${}^{156}$Gd, given as a function of the $\gamma$-ray 
         energy, used in the SLO approach.}
    \label{fig:hfbpsf}
  \end{center}
\end{figure}

% T(E_\gamma)= 2\pi E_\gamma^3\sum_{i=1}^2{\frac{\sigma_i E_\gamma \Gamma_i^2}{(E_\gamma^2 - E_i^2)^2 + E_\gamma^2\Gamma_i^2}}

% $$
%\begin{equation}
%T_{E1}(E_\gamma)= 2\pi E_\gamma^3{\frac{1}{3\pi(\hbar c)^2}}\sum_{i=1}^4{\frac{\sigma_i E_\gamma \Gamma_i^2}{(E_\gamma^2 - E_i^2)^2 + E_\gamma^2\Gamma_i^2}},\  
%\end{equation}

\begin{equation}
  P(E_a,E_b) = \frac{\mathrm{d}P}{\mathrm{d}E}(E_a,E_b) \, \delta E = \frac{\rho(E_b) T_{E1}(E_\gamma)}{\int_{0}^{E_a} \rho(E'_b) T_{E1}(E'_\gamma)  \, \mathrm{dE}'_b} \, \delta E \, , \quad  E'_\gamma = E_a - E'_b\, ,
\label{eq:continuum_prob}
\end{equation}
where $\delta E$ is a finite energy step in our computations.  $T_{E1}(E_\gamma)$  refers to the E1 Photon Strength Function $f_{E1}(E_\gamma)$ (PSF) depending on cross section ($\sigma_i$),  the width ($\Gamma_i$)  and  energy ($E_i$) of the resonances. It is written as 
\begin{eqnarray}
 T_{E1}(E_\gamma) = 2 \pi \, E^3_\gamma \, f_{E1}(E_\gamma) , \  \text{and}\,  \nonumber \\
\ f_{E1}(E_\gamma)= {\frac{1}{3(\pi \hbar c)^2}}\sum_{i=1}^4{\frac{\sigma_i E_\gamma \Gamma_i^2}{(E_\gamma^2 - E_i^2)^2 + E_\gamma^2\Gamma_i^2}},\  
\end{eqnarray}
where values of E$_i$, $\sigma_i$ and width $\Gamma_i$ are mentioned in Table~\ref{tab:PSFParameters}  and  $\rho (E_b)$ is the nuclear level density (NLD). 
We note that we add the  two small (pigmy) E1 resonances of the same Lorentzian type ($i=3,4$) to the PSF in Eq. (3) in order to check the effect of those two resonances on the $\gamma$-ray spectrum~\cite{Kopecky90, Capote2009:RIPL3, Shibata2011:JENDL4}, while we used  only the first two major E1 resonances in the previous publication~\cite{Hagiwara:2018kmr}. Since these four resonances are all E1-type, we can construct  the probability tables according to Eq. (2) to generate the $\gamma$-ray spectrum~\footnote{The effect of including the two additional resonances on the gross $\gamma$-ray spectrum was not so significant as the case with only the two major E1 resonances. While we add the two E1 resonances ($i=3,4$), we do not add the M1 and E2 resonances to the PSF in Eq. (3). If we included  the M1 and E2 resonances in Eq. (3), we would have to separate the NLD of Eq. (2)  into the NLDs for positive-parity and negative-parity levels  and we have not done it within this paper.}. The corresponding NLD~\cite{Capote2009:RIPL3, Goriely2007:HFB, Goriely2008:HFB}  and the PSF~\cite{Shibata2011:JENDL4} used for $^{\rm 156}$Gd are shown  in Fig.~\ref{fig:hfbpsf} (left and right respectively). The recent review on the NLD and the PSF can be found in Ref.~\cite{Capote2009:RIPL3, Goriely2019}.

%__________________________________________________________________________________________________%
% TABLE BEGIN
\begin{table}[h]
    \centering
    \caption{Parameter values for the PSF of the ${}^{156}$Gd 
             nucleus~\cite{Shibata2011:JENDL4}. We use the E1 resonances only for our model.}
%              We use only the first and socond resonances for our model.} 
    \begin{tabular}{rccc}
        \hline
        Index $i$ & Cross-section $\sigma_i$ & Energy $E_i$ & Width $\Gamma_i$ \\
                  & [mb]                     & [MeV]        & [MeV]\\
        \hline
       (E1) 1         & 242                     & 15.2        & 3.6 \\
       (E1) 2         & 180                     & 11.2        & 2.6 \\
       (E1) 3         & 2.0                     & 6.0        & 2.0 \\
       (E1) 4         & 0.4                     & 3.0        & 1.0 \\
       (M1) 5         & 2.03                     & 7.62        & 4.0 \\
       (E2) 6         & 3.69                     & 11.7        & 4.24 \\
        \hline
    \end{tabular}
    \label{tab:PSFParameters}
\end{table}
% TABLE END
%__________________________________________________________________________________________________%

%__________________________________________________________________________________________________%
% FIGURE BEGIN
% FIGURE END
%__________________________________________________________________________________________________%

\section{Final model performance}
\begin{figure}[h]
  \begin{center}
   \includegraphics[width=0.5\textwidth]{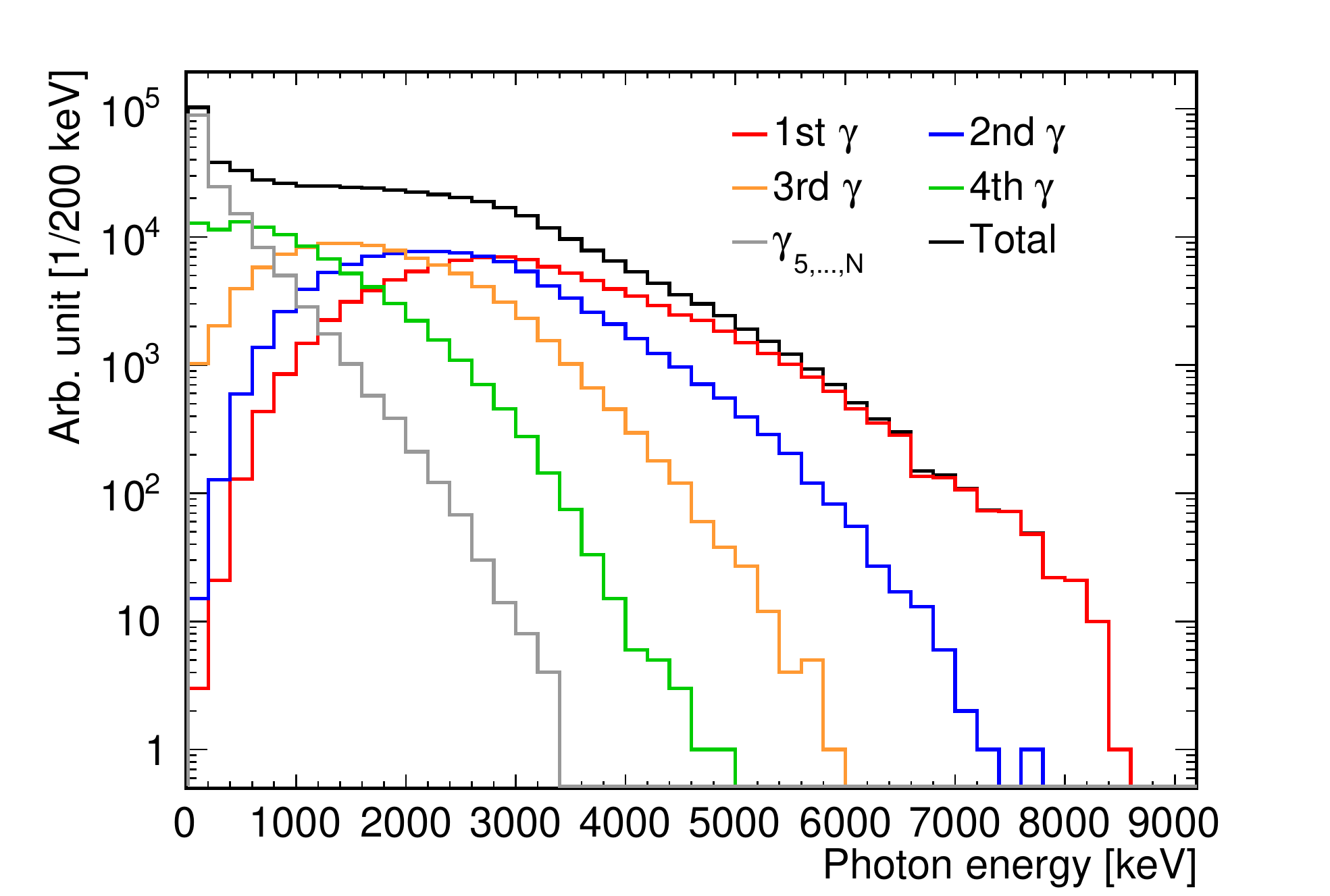}
     \caption{ Continuum component (black) according to our model for the 
           $\gamma$-ray energy spectrum from the ${}^{155}$Gd$(n,\gamma){}$ reaction 
           and its composition in terms of contributions 
         from the first (red), second (blue), third (orange), fourth (green) $\gamma$ ray and
         other $\gamma$ rays (gray).  }
    \label{fig:155gammaMC}
  \end{center}
\end{figure}

\begin{figure}[h]
  \begin{center}
    \includegraphics[width=0.8\textwidth]{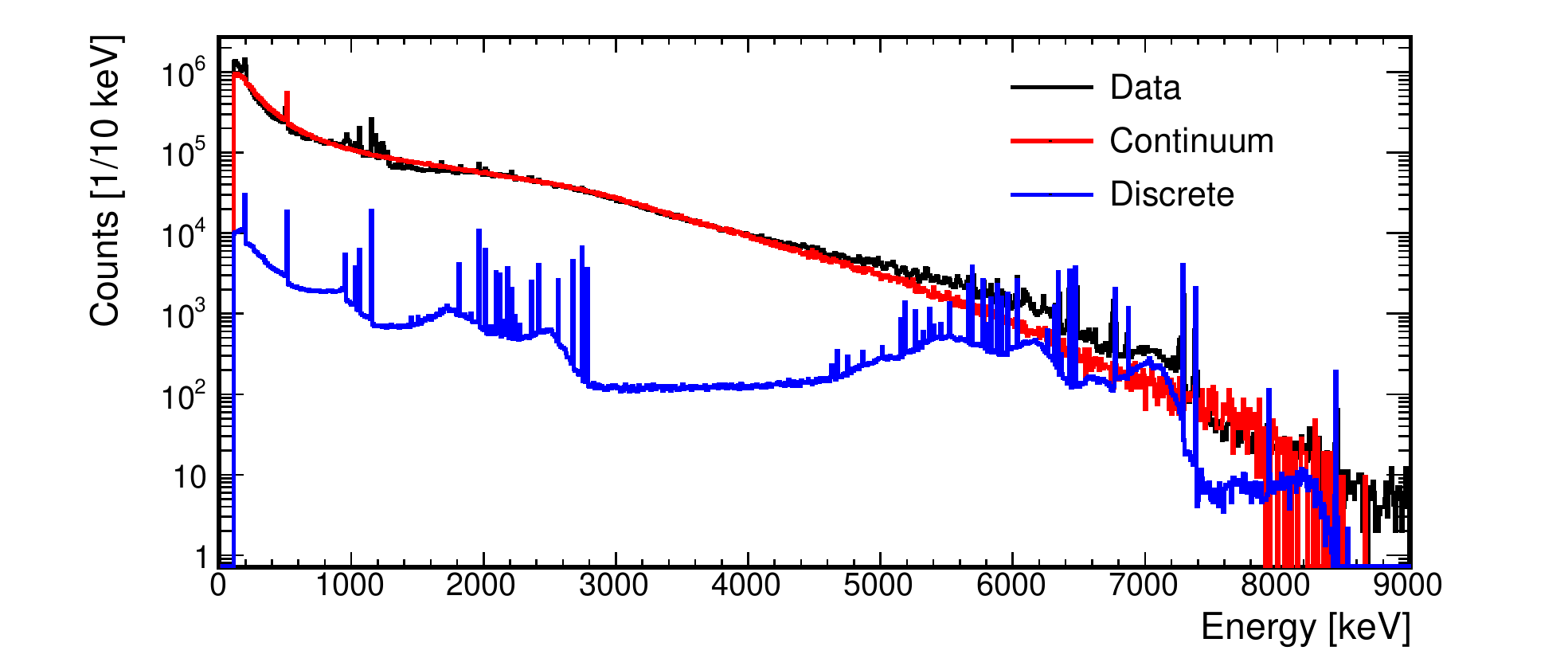}
    \caption{ The continuum and the discrete components of the spectrum generated by the MC shown separately along with the data spectrum.}
    \label{fig:155disccont}
  \end{center}
\end{figure}

\begin{figure}[h]
  \begin{center}
    \includegraphics[trim=0.1cm 0.1cm 0.5cm 0.1cm, clip=true,width=0.49\textwidth] 
    {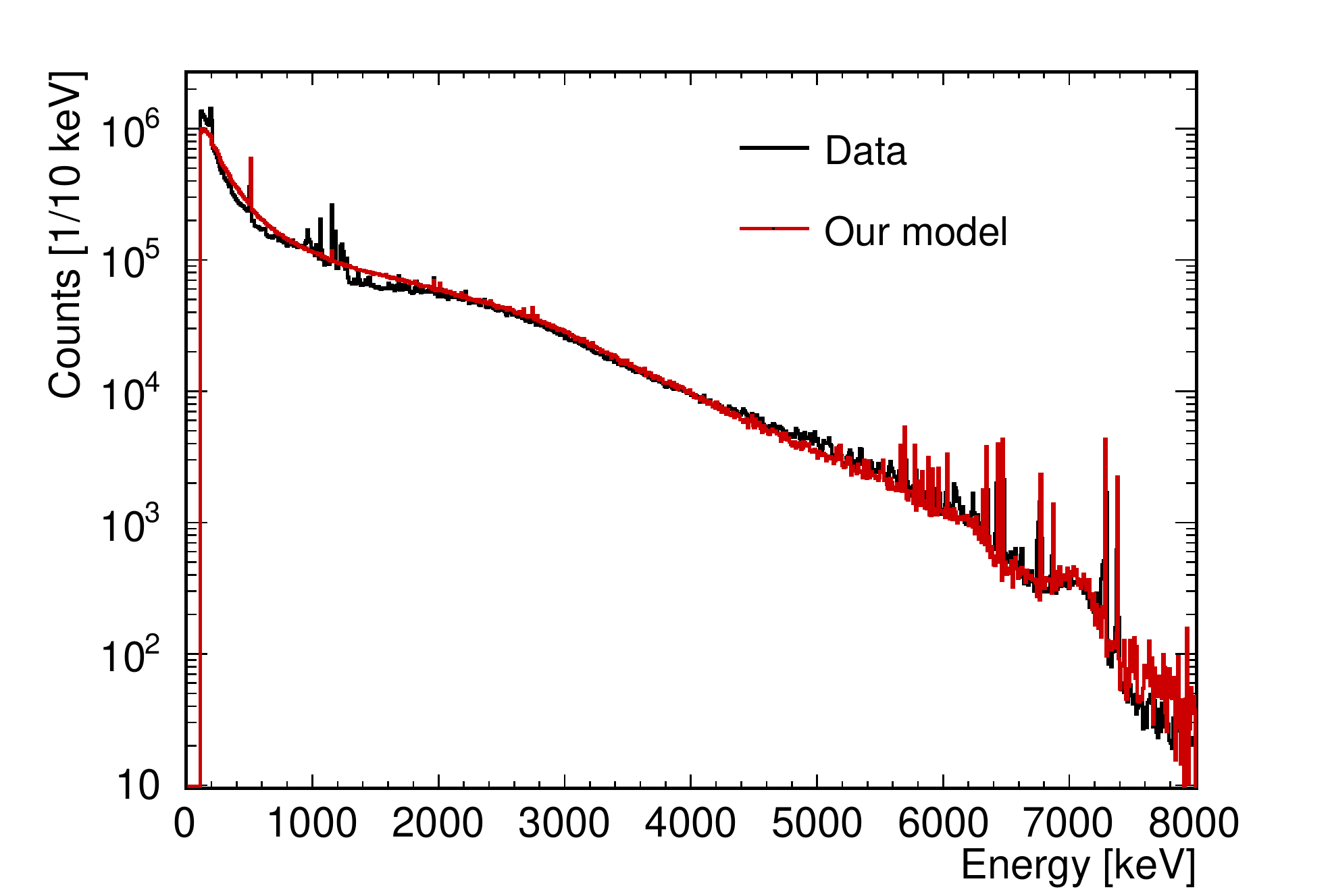}
     \includegraphics[trim=0.1cm 0.1cm 0.5cm 0.1cm, clip=true,width=0.49\textwidth] 
    {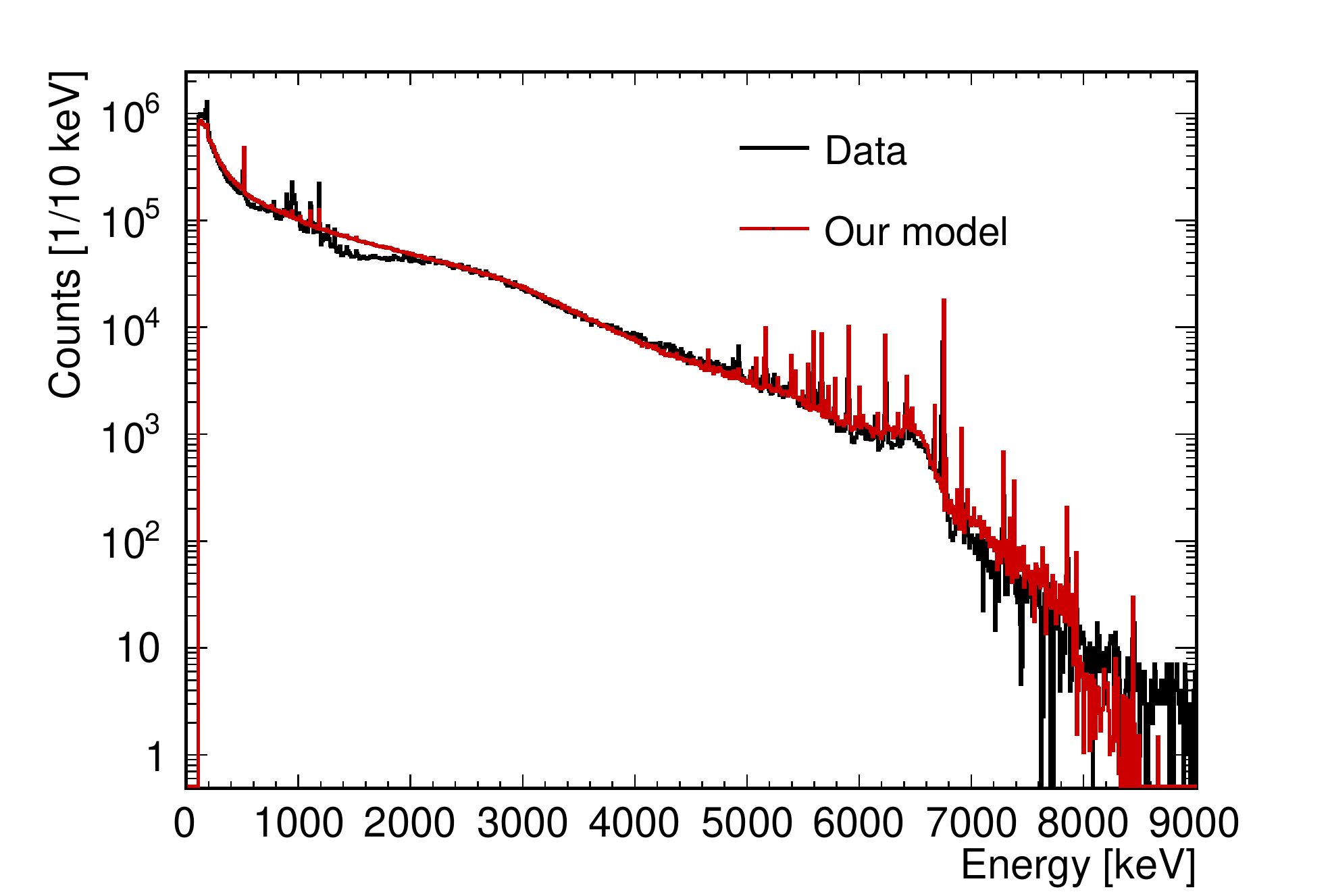}
   \caption{Single Energy spectrum (M1H1) generated by our model compared with the data for $^{155}$Gd on left, and $^{\rm nat}$Gd on right.}
    \label{fig:datamc155}
  \end{center}
\end{figure}

We first generate the continuum part of the $\gamma$-ray spectrum  in $^{\rm 156}$Gd according to Eq.\ (2).  The result is  shown in Fig.~\ref{fig:155gammaMC}. We then generate the discrete part according to the relative intensities listed in Table 3 and then  compare these two parts with the observed spectrum. We determine the fraction of the discrete part in the total number of events  to be 
 2.78$\pm$0.02\% of the data above 0.11 MeV.  The remaining dominant contribution of 97.22$\pm$0.02\% comes from the continuum part of the energy levels in $^{\rm 156}$Gd. The continuum and the discrete components generated by our MC model are shown separately here for $^{155}$Gd, along with the data in Fig.~\ref{fig:155disccont}. They are added in the corresponding fractions  in Fig.~\ref{fig:datamc155}-left. The data spectrum matches well with our MC spectrum.
% Spectral components of the discrete part are added 
%and are tuned with that of $^{155}$Gd data. 

% \textcolor{red}{stats:\#number ??}

The  MC generated spectrum for $^{\rm nat}$Gd(n, $\gamma$) should naturally comprise  the spectrum for $^{\rm 155}$Gd(n, $\gamma$) and $^{\rm 157}$Gd(n, $\gamma$), as is obvious with the data spectra in Fig.~\ref{fig:datacomp}. So, the spectrum for $^{\rm nat}$Gd(n, $\gamma$) is obtained by adding the MC spectra  generated for $^{\rm 155}$Gd(n, $\gamma$) and $^{\rm 157}$Gd(n, $\gamma$) in the required ratio of
their relative cross-section and abundance, as is shown in Fig.~\ref{fig:datamc155}-right. 
%__________________________________________________________________________________________________%
% FIGURE BEGIN
% FIGURE END
%__________________________________________________________________________________________________%
%__________________________________________________________________________________________________%
% FIGURE BEGIN
% FIGURE END
%__________________________________________________________________________________________________%
 
%__________________________________________________________________________________________________%
% FIGURE BEGIN
% FIGURE END
%__________________________________________________________________________________________________%

The spectra shown above are single  energy spectra  (M1H1), which constitute the most dominant ($\sim$70\%) fraction of the data. 
% (figure~\ref{fig:dtmcstat}-left). 
In fact,  good agreement is found between all the MC generated spectra and the subsamples of data  for different observed multiplicities M.
% , as in figure~\ref{fig:dtmcstat}-right. 
Exemplarily, the M2H2 and M3H3 spectra are shown in Appendix~\ref{sec:Multigamma}.
\section{Conclusion}
\label{sec:results}
%==================================================================================================%

\begin{figure}[H]
  \begin{center}
     \includegraphics[trim=0.1cm 0.1cm 0.5cm 0.1cm, clip=true,width=0.8\textwidth] 
    {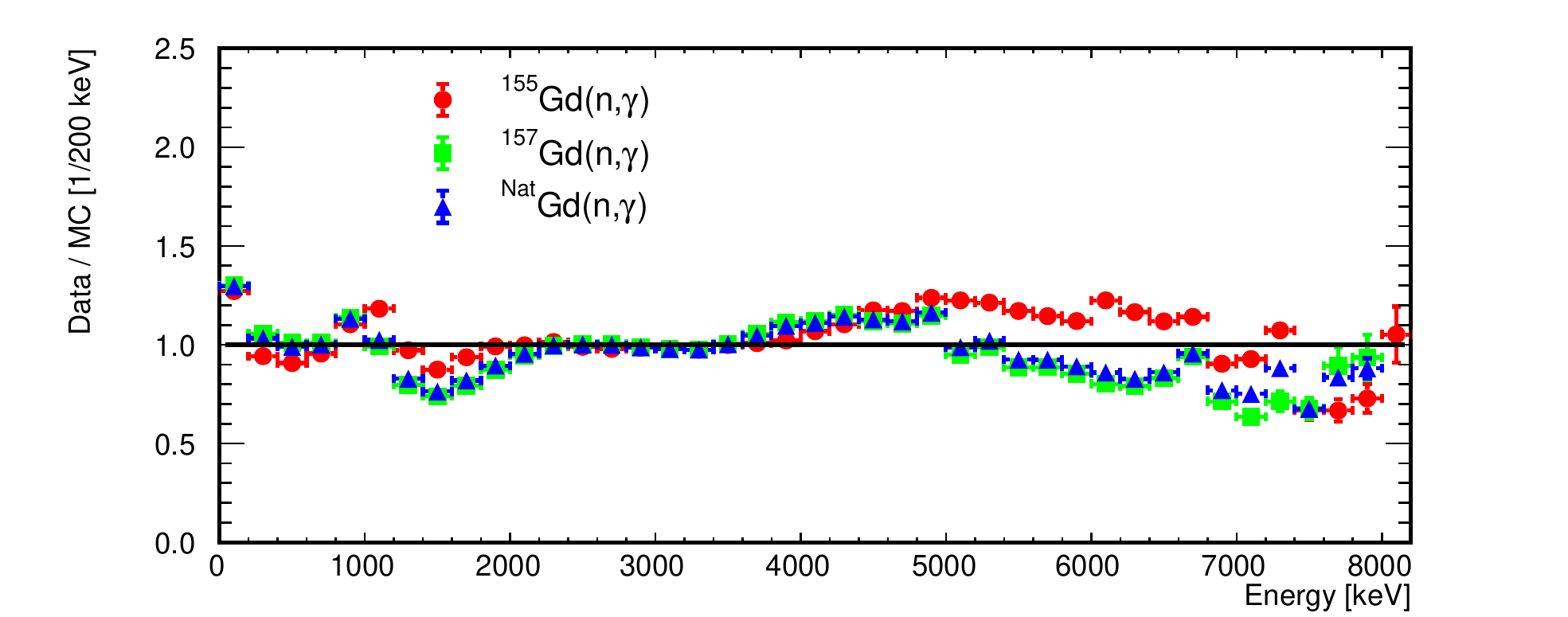}
    \caption{Ratio of data by MC for the single $\gamma$ events (M1H1 + M1H2) obtained for $^{\rm 157}$Gd(n,$\gamma$), $^{\rm 155}$Gd(n,$\gamma$) and $^{\rm nat}$Gd(n,$\gamma$) cases.}
    \label{fig:gdfinalratio}
  \end{center}
\end{figure}

The $\gamma$-ray  spectra generated by our ANNRI-Gd model agree not only with the individual $^{\rm 155}$Gd and $^{\rm 157}$Gd data set, but also with $^{\rm nat}$Gd data set, which are entirely independent\footnote{The data of $^{\rm 155}$Gd and $^{\rm 157}$Gd were used to tune the discrete part of our MC model, while the $^{\rm nat}$Gd data was untouched during the building of our MC.}. 
We show the ratio of data/MC in bins of 200 keV for  $^{\rm 155}$Gd, $^{\rm 157}$Gd, and $^{\rm nat}$Gd in Fig.~\ref{fig:gdfinalratio}, for the single $\gamma$-ray $M=1$ events as an approximate representation of the goodness of our model.
 For the presented single $\gamma$-ray spectrum with 
the 200 keV binning, the mean deviation of the single ratios from the mean ratio is about 17\% for each of $^{\rm 157}$Gd,  $^{\rm 155}$Gd and $^{\rm nat}$Gd spectra.
 The same ratios for the $M=2$ and  $M=3$ samples are shown in Fig.~\ref{fig:gdfinalratio23}. They are all in good agreement at a similar level to  those published for the $^{\rm 157}$Gd(n, $\gamma$) reaction \cite{Hagiwara:2018kmr}. With this article, we have completed a consistent model (ANNRI-Gd Model) to generate the gross spectrum for the thermal $^{\rm 155}$Gd,  $^{\rm 157}$Gd and $^{\rm nat}$Gd(n, $\gamma$) reaction. 
 
In comparison,  the more sophisticated model~\cite{Valenta2015:GdTwoStepGamCasc} tries to include a small contribution of M1 (scissors mode) or E2 resonance around 3 MeV to PSF in order to explain the energy spectra in the sample of two-step cascade $\gamma$ rays from the thermal neutron capture reactions. The DANCE experiment~\cite{Baramsai2013:DANCE, Chyzh:DANCE} also suggested a need of small resonances (M1 or E2) around 3 MeV in addition to the major E1 PSFs in order to explain the $\gamma$-ray energy spectra of the multiplicity $M$=2, though the data of the $^{\rm 155, 157}$Gd($n, \gamma$) reactions were taken in the neutron kinetic energies in the tens of eV.  
To further refine the present modeling, we intend to work on a sample of 2$\gamma$ rays including strong discrete cascade transitions. 
We note that those samples constitute a few \% of the total number of capture events. 
As these previous articles point out, we must handle the positive-parity states and negative-parity states separately in the NLD or in any discrete levels in order to take into account the E1 transition or M1/E2 transition correctly during the cascade.

  After we submitted this article in August, 2019,  the Daya Bay Collaboration, one of the most advanced reactor-neutrino experiments, has reported a Monte Carlo study of the $\gamma$-ray spectra from the thermal neutron capture on $^{155}$Gd and $^{157}$Gd  and shown the large discrepancies in the $\gamma$-ray spectra generated by various Monte Carlo models~\cite{DayaBay2}. 
We compare our spectrum with their result in the Appendix B. It shows clearly that our data and our MC model will help resolve such discrepancies in the gross $\gamma$-ray spectrum generated by various MC models for  the thermal $^{\rm 155}$Gd,  $^{\rm 157}$Gd and $^{\rm nat}$Gd(n, $\gamma$) reactions. 
% The {\bf ANNRI-Gd} model very well agrees with the data taken with ${}^{157}$Gd, ${}^{155}$Gd and also  ${}^{\rm Natural}$Gd.%, within ?? \% of max. deviation.
%  energy spectrum deviation of $\sim$ 15\% at 200keV binning.
%  Discrete component of ${}^{155}$Gd: $\sim$3\% and for ${}^{157}$Gd: $\sim$7\%, so a finer tuning for the latter may add further accuracy to our model.
%   

%__________________________________________________________________________________________________%
% FIGURE BEGIN
% FIGURE END
%__________________________________________________________________________________________________%
% % % % 
% % % % \begin{figure}[H]
% % % %   \begin{center}
% % % %   \includegraphics[width=0.49\textwidth]{nat_hist_MH22_c6resize.pdf}
% % % %   \includegraphics[width=0.49\textwidth]{nat_hist_MH33_c6resize.pdf}
% % % %     \caption{M2H2 on left and M3H3 on right.}
% % % %     \label{fig:gdfinalspec23}
% % % %   \end{center}
% % % % \end{figure}

%==================================================================================================%
\section*{Acknowledgement}
\label{sec:Acknowledgements}
%==================================================================================================%

This work is supported by the JSPS Grant-in-Aid for Scientific Research on Innovative Areas 
(Research in a proposed research area) No. 26104006 and No. 15K21747. It benefited from the use of the 
neutron beam of the JSNS and the ANNRI detector at the Material and Life Science Experimental 
Facility of the Japan Proton Accelerator Research Complex.

\appendix{
\section*{Appendices}
\section{Double/Triple $\gamma$-Ray Spectra}
\label{sec:Multigamma}

Apart from the single $\gamma$-ray events (M=1), the M=2 and M=3 $\gamma$-ray events are also observed. 
The M1H1 sample is the most dominant one, followed by the M1H2 sample. Our model agrees with data in both cases. 
% \textcolor{red}
The relative fraction (in \%)  in the Data and MC for the different subsamples (M1H1, M1H2, M2H2 etc) are shown in Fig. A1. Data and MC agree well. The corresponding spectra for the M2H2 and M3H3  samples generated by our model also agree well with the $^{\rm 155}$Gd(n,$\gamma$) and the $^{\rm nat}$Gd(n,$\gamma$) data, as shown in Fig. A2 and Fig. A3, respectively.  The ratios of data/MC  are also shown in Fig. A4.

% With increased multiplicity, the slope of the spectrum grows softer, owing to reduced probability of emitting higher energy gamma rays, as also seen in Fig.~\ref{fig:multispectra} and independent checks from simulations as in the Fig.~\ref{fig:Gd158_PDD}.
%__________________________________________________________________________________________________%
% FIGURE BEGIN
\begin{figure}[H]
  \begin{center}
    \includegraphics[trim=0.1cm 0.1cm 0.5cm 0.1cm, clip=true,width=0.65\textwidth] 
    {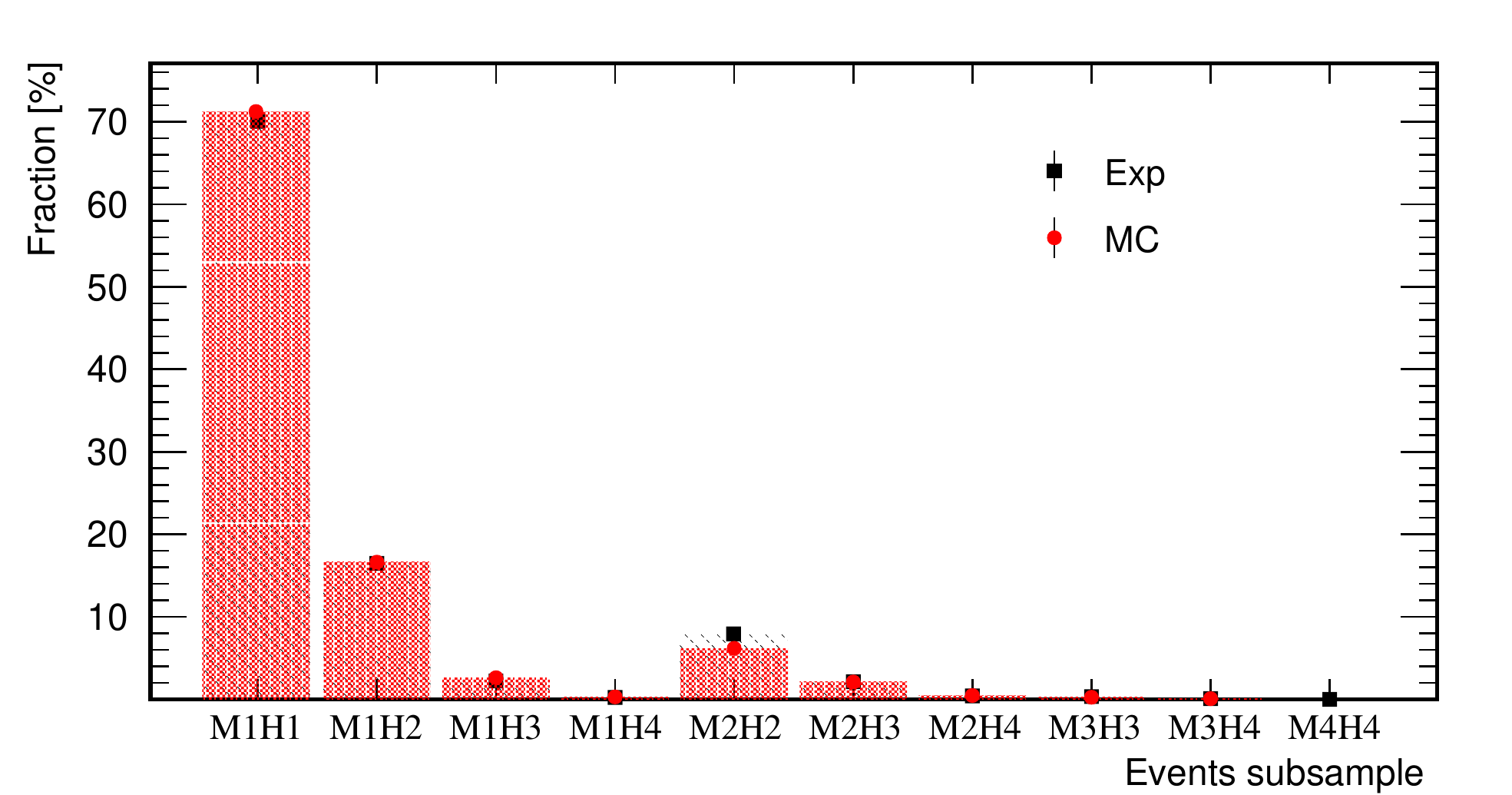}
    \caption{Relative fraction in \% in the Data and MC for the different subsamples: M1H1, M1H2, M2H2 etc.
% Left: Relative counts/statistics in \%. 
%    Right: Ratio of data and MC. \textcolor{red}{X-axis same as left; may change after checking with threshold 500keV: Tanaka}
}
    \label{fig:datamcclass}
  \end{center}
\end{figure}
% FIGURE END
%__________________________________________________________________________________________________%

%__________________________________________________________________________________________________%
% FIGURE BEGIN
\begin{figure}[H]
  \begin{center}
    \includegraphics[trim=0.1cm 0.1cm 0.5cm 0.1cm, clip=true,width=0.49\textwidth] 
    {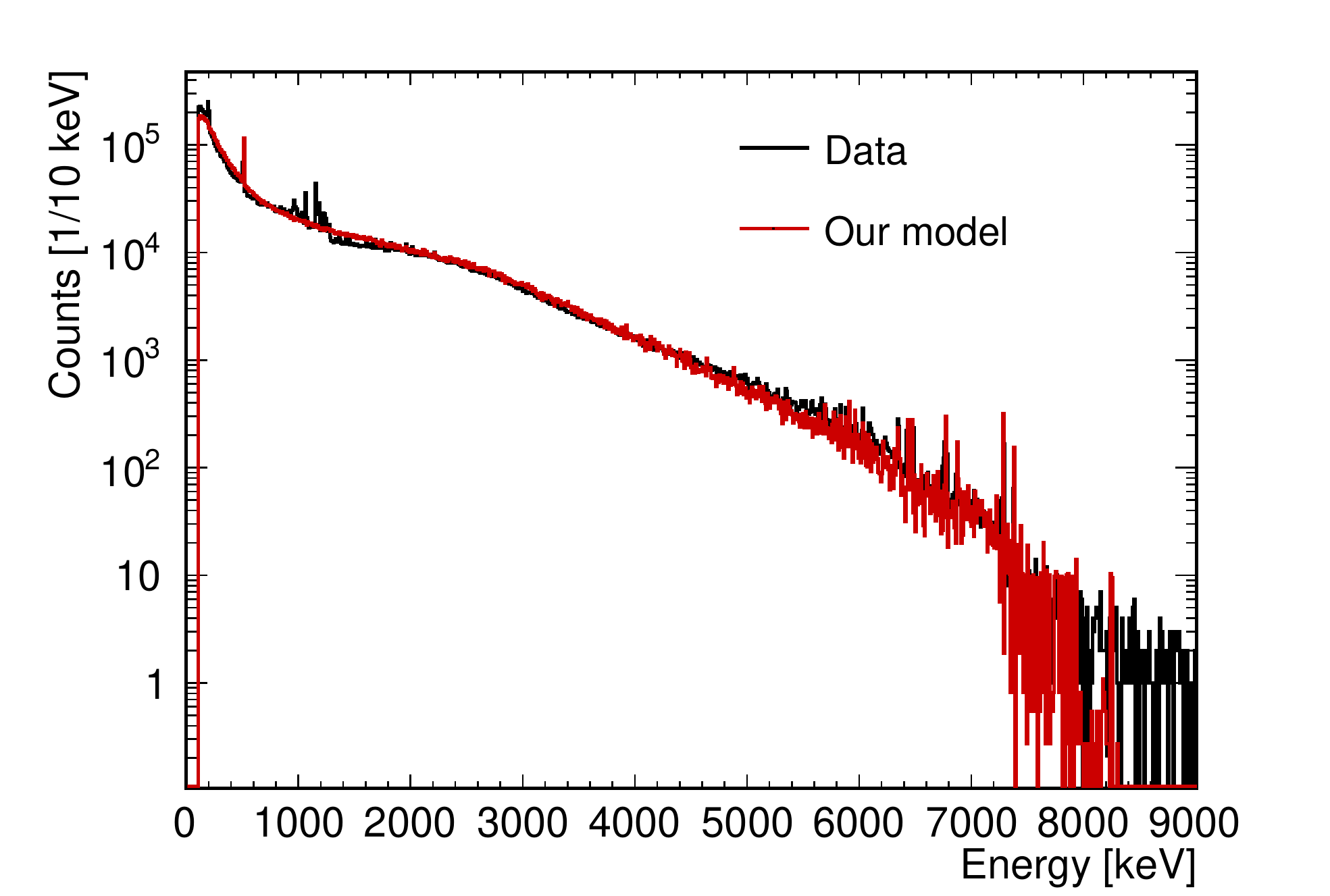}
     \includegraphics[trim=0.1cm 0.1cm 0.5cm 0.1cm, clip=true,width=0.49\textwidth] 
    {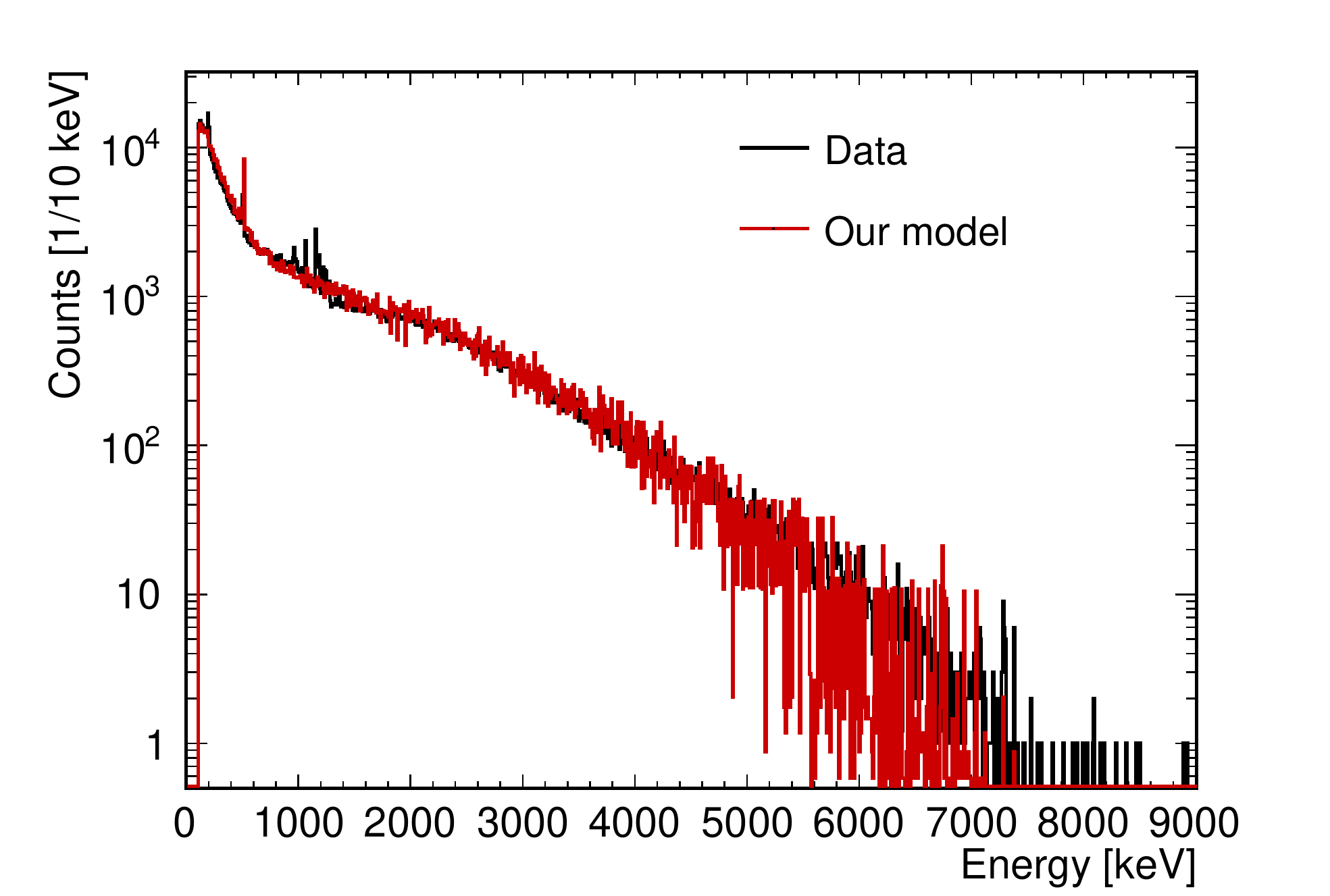}
   \caption{The $^{\rm 155}$Gd(n,$\gamma$) spectra for M2H2 (left) and M3H3 (right) samples from data and our model MC.}
    \label{fig:datamc155mh23}
  \end{center}
\end{figure}
% FIGURE END
%__________________________________________________________________________________________________%

%__________________________________________________________________________________________________%
% FIGURE BEGIN
\begin{figure}[H]
  \begin{center}
    \includegraphics[trim=0.1cm 0.1cm 0.5cm 0.1cm, clip=true,width=0.49\textwidth] 
    {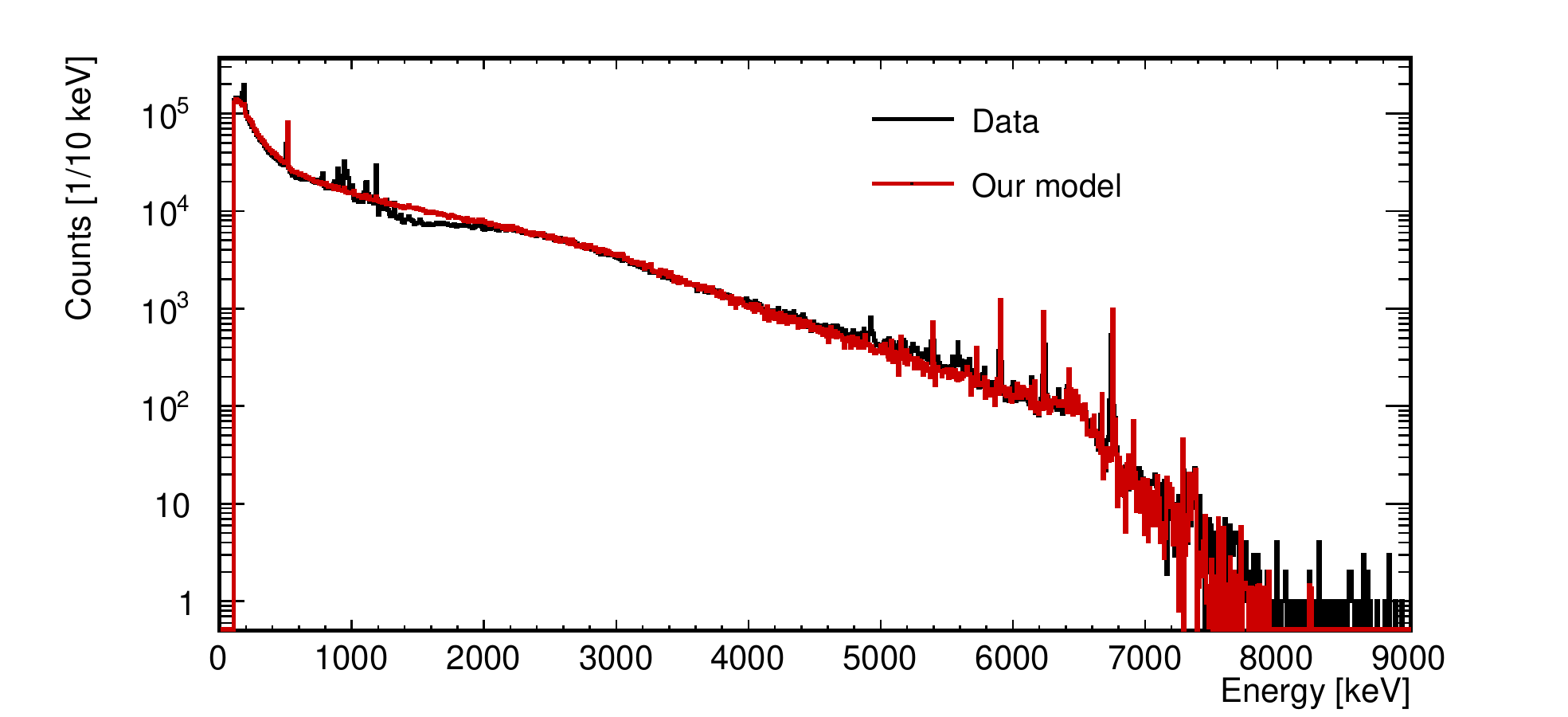}
     \includegraphics[trim=0.1cm 0.1cm 0.5cm 0.1cm, clip=true,width=0.49\textwidth] 
    {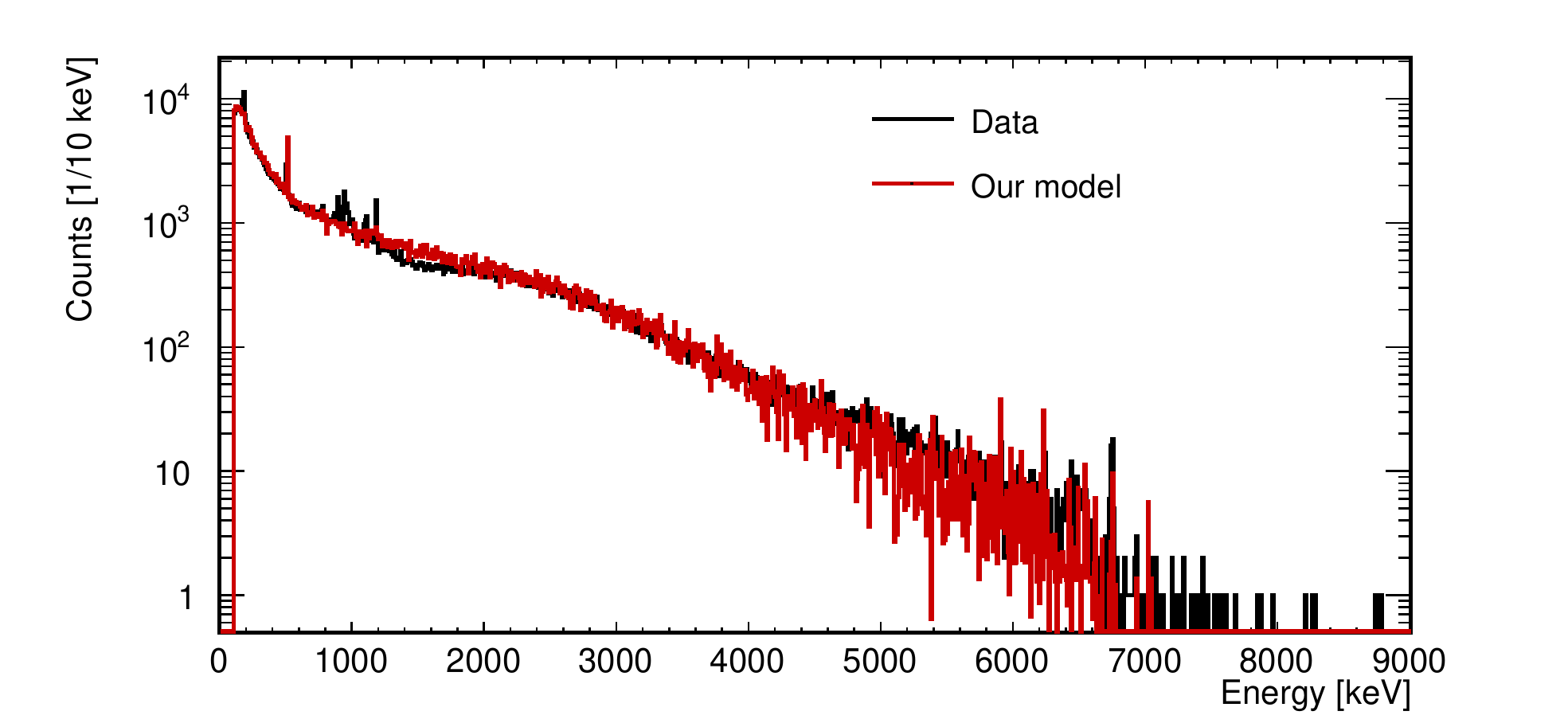}
   \caption{The $^{\rm nat}$Gd(n,$\gamma$) spectra for M2H2 (left) and M3H3 (right) samples from data and our model MC.}
    \label{fig:datamcnatmh23}
  \end{center}
\end{figure}
% FIGURE END
%__________________________________________________________________________________________________%

\begin{figure}[H]
  \begin{center}
  \includegraphics[width=0.49\textwidth]{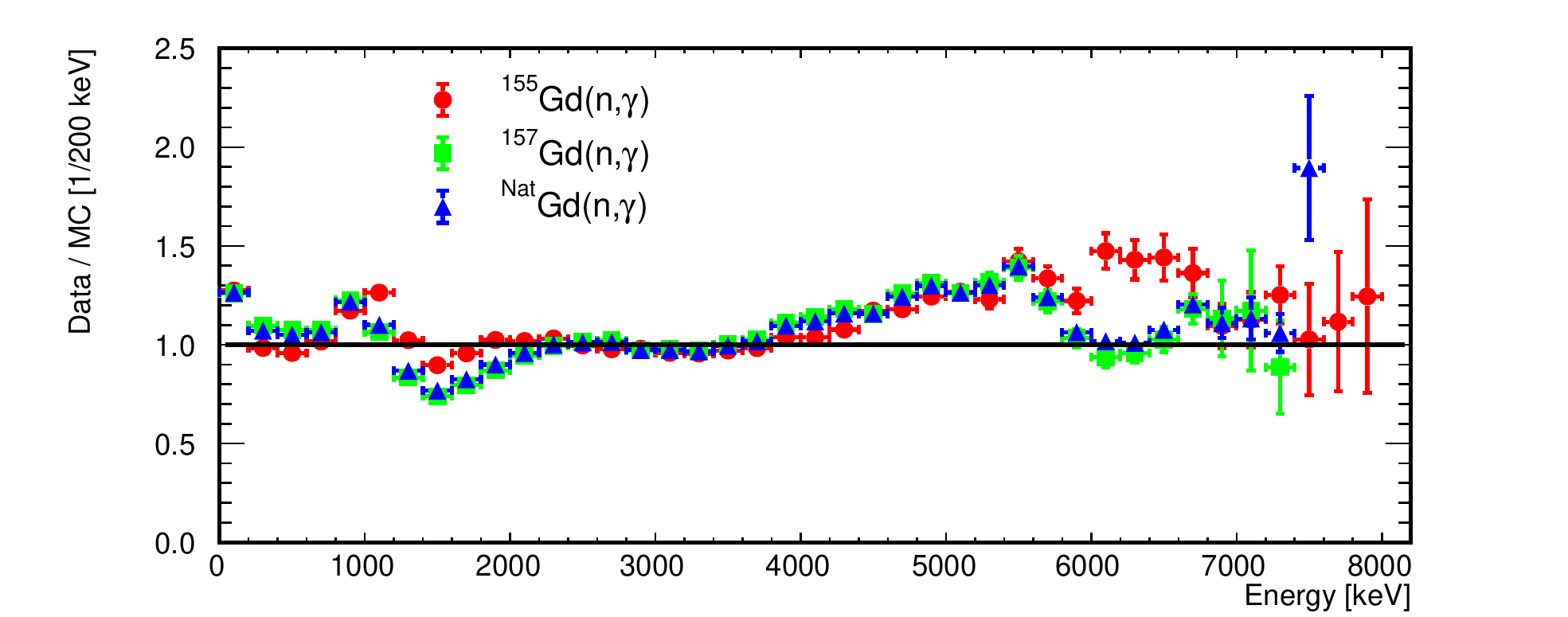}
  \includegraphics[width=0.49\textwidth]{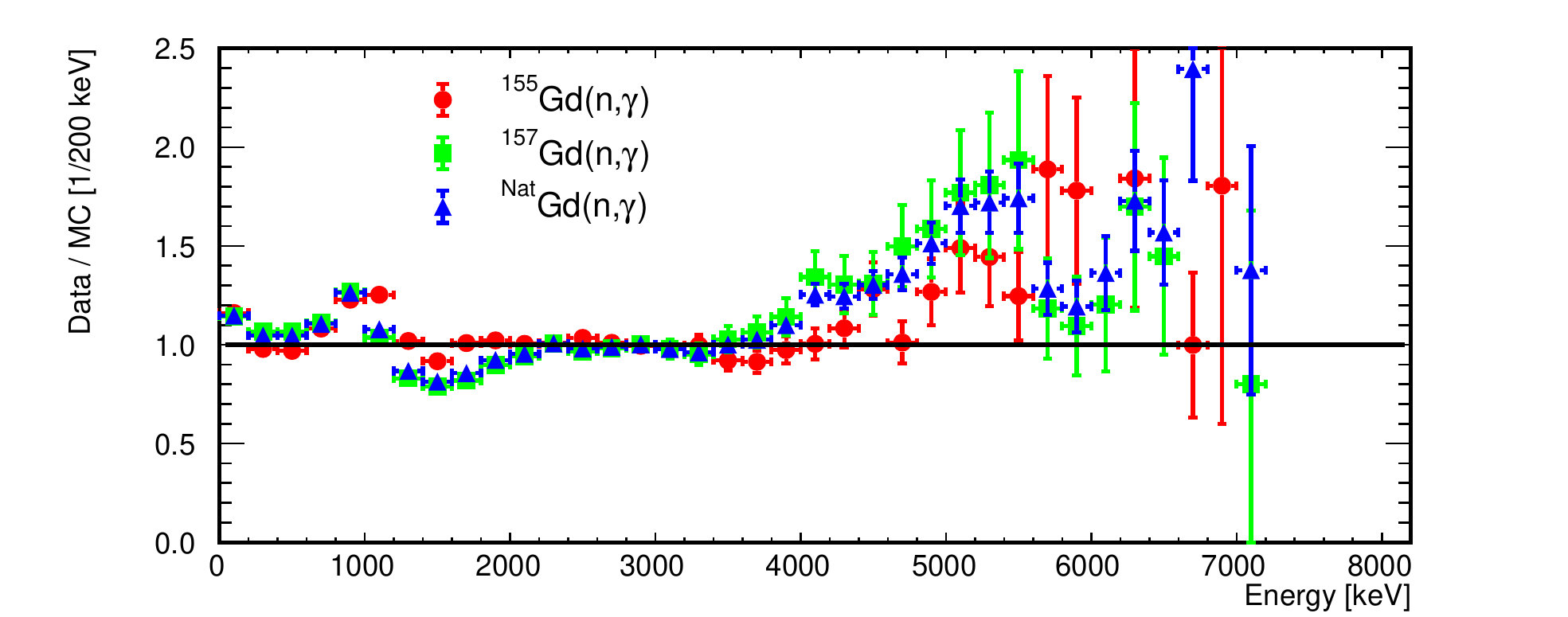}
    \caption{Ratios of data/MC for M2H2 (left) and  M3H3 (right) samples obtained for $^{\rm 157}$Gd(n,$\gamma$), $^{\rm 155}$Gd(n,$\gamma$) and $^{\rm nat}$Gd(n,$\gamma$) cases.}
    \label{fig:gdfinalratio23}
  \end{center}
\end{figure}

\section{Comprison of ANNRI-Gd model with Various Models Reported by Daya Bay Collaboration}

     Recently, the Daya Bay Collaboration showed the $\gamma$-ray  spectra of  the thermal neutron capture on $^{\rm 155}$Gd and  $^{\rm 157}$Gd  in Figs. 5(a) and (b) of Ref.~\cite{DayaBay2}, which are produced by various MC models. We use their Fig.5(a) of Ref.~\cite{DayaBay2}, which they quote as the energy distribution of the deexcitation gammas of $^{\rm 155}$Gd. 
We add to their Fig. 5(a)  a single $\gamma$-ray spectrum (purple line) as shown in Fig. B1, which is generated by our ANNRI-Gd model. We note that  we are comparing the shape among various models and that we normalize the histogram by the total entries in the figure~\footnote{We read the entries off the histogram of Fig. 5(a) of Ref.~\cite{DayaBay2} and make Fig. B1. Thus, the values of the histogram may be
different from the original values by a few \%.}. 
 We showed already in Figs.4 and 10 that our predictions agree with our measured spectra within about  17\% at 200 keV binning. Our model agrees well with the Model 1 (a native Geant4 model) for $E_\gamma $ above 2 MeV, but our model disagrees with the Model 1 below 2 MeV. Other Models generate the spectra which are very different from ours in shape. We would like to stress again that we can discuss the small structure at 2-3 MeV such as scissors mode only after we understand the gross spectrum over the entire energy region.

\begin{figure}[H]
  \begin{center}
    \includegraphics[trim=0.1cm 0.1cm 0.5cm 0.1cm, clip=true,width=0.65\textwidth] 
    {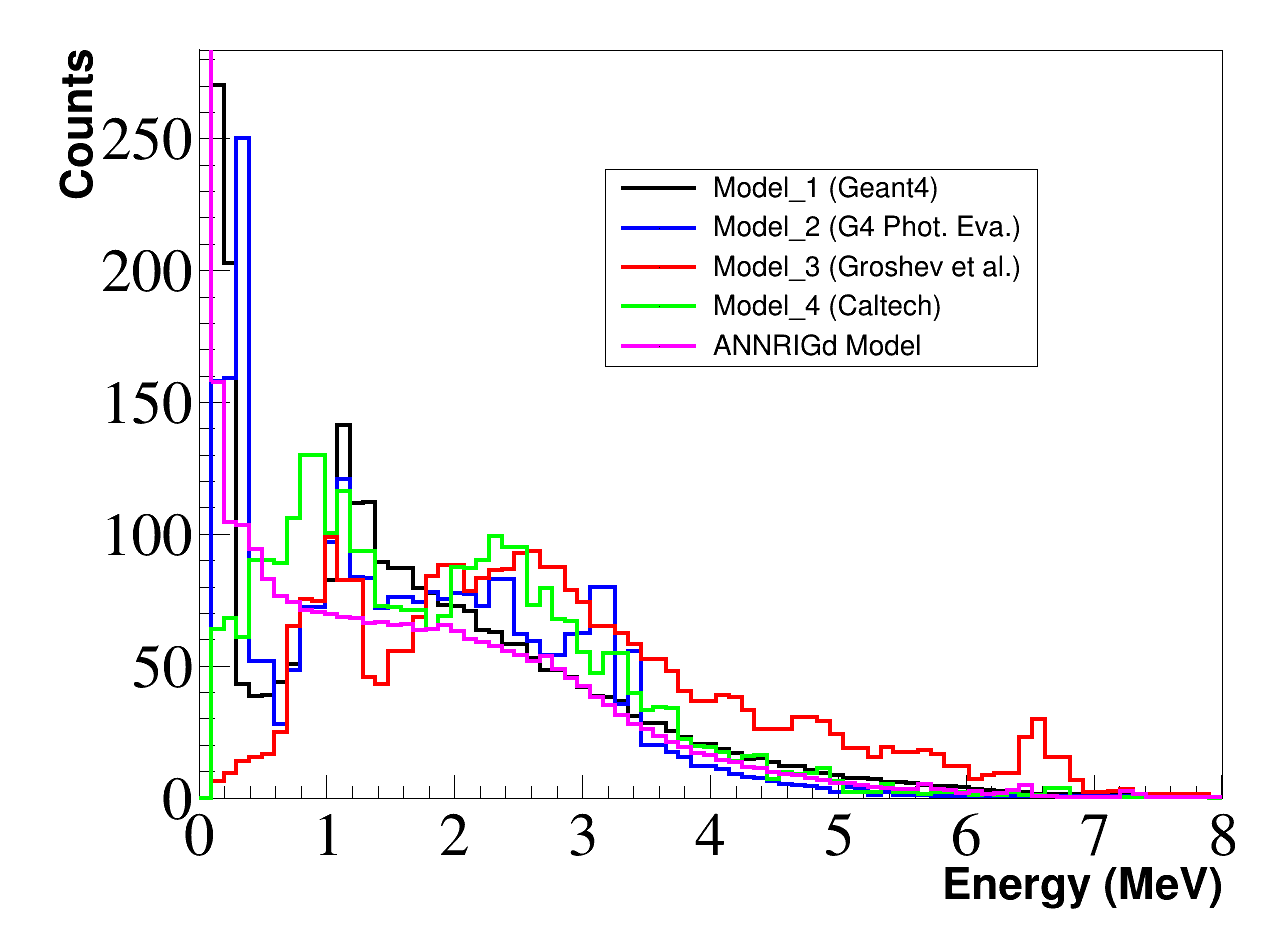}
    \caption{Energy distribution of  $\gamma$ rays from the $^{\rm 155}$Gd(n,$\gamma$) reaction for the various Models 1-4 shown in Ref.~\cite{DayaBay2} and our ANNRI-Gd model prediction (purple line). 
}
    \label{fig:datamcclass}
  \end{center}
\end{figure}

\end{document}